\documentclass[12pt,letterpaper]{article}
\usepackage{epsfig,rotating,setspace,latexsym,amsmath,epsf,amssymb,amsfonts,bm,theorem,cite,enumerate,longtable,accents,float,physics}
\usepackage{algorithm,graphicx,epsf,authblk,epstopdf,url,xcolor, soul,multirow,bbm}
\usepackage{mathtools,comment}
\usepackage[center]{qtree}
\usepackage{tree-dvips}
\usepackage[linguistics]{forest }
\usepackage[noend]{algpseudocode}
\usepackage{hyperref}
\usepackage{dsfont}

\newtheorem{theorem}{Theorem}

\newtheorem{definition}{Definition}
\newtheorem{remark}{Remark}
\newtheorem{lemma}{Lemma}

\newenvironment{Proof}[1]{\medskip\par\noindent{\bf Proof:\,}\,#1}{{\mbox{\,$\blacksquare$}\par}}

\DeclareMathOperator{\diag}{diag}
\newcommand{\prob}[2]{\mathbb{P}_{#1}\left(#2 \right)}

\allowdisplaybreaks
\date{}

\title{What If, But Privately: \\ Private Counterfactual Retrieval\footnote{A part of this paper was presented at IEEE ISIT 2025.}}

\author{Shreya Meel \qquad Mohamed Nomeir \qquad Pasan Dissanayake\newline
 Sanghamitra Dutta \qquad Sennur Ulukus \\
\normalsize Department of Electrical and Computer Engineering \\
\normalsize University of Maryland, College Park, MD 20742 \\
\normalsize {\qquad {\it smeel@umd.edu} \qquad \it mnomeir@umd.edu} \qquad {\it pasand@umd.edu}  \newline {\it sanghamd@umd.edu} \qquad {\it ulukus@umd.edu}}
 
\begin{document}

\maketitle

\begin{abstract}
    Transparency and explainability are two important aspects to be considered when employing black-box machine learning models in high-stake applications. Providing counterfactual explanations is one way of catering this requirement. However, this also poses a threat to the privacy of the institution that is providing the explanation, as well as the user who is requesting it. In this work, we are primarily concerned with the user's privacy who wants to retrieve a counterfactual instance, without revealing their feature vector to the institution. Our framework retrieves the exact nearest neighbor counterfactual explanation from a database of accepted points while achieving perfect, information-theoretic, privacy for the user. First, we introduce the problem of private counterfactual retrieval (PCR) and propose a baseline PCR scheme that keeps the user's feature vector information-theoretically private from the institution. Building on this, we propose two other schemes that reduce the amount of information leaked about the institution database to the user, compared to the baseline scheme. Second, we relax the assumption of mutability of all features, and consider the setting of immutable PCR (I-PCR). Here, the user retrieves the nearest counterfactual without altering a private subset of their features, which constitutes the immutable set, while keeping their feature vector and immutable set private from the institution. For this, we propose two schemes that preserve the user's privacy information-theoretically, but ensure varying degrees of database privacy. Third, we extend our PCR and I-PCR schemes to incorporate user's preference on transforming their attributes, so that a more actionable explanation can be received. Finally, we present numerical results to support our theoretical findings, and compare the database leakage of the proposed schemes.
\end{abstract}

\section{Introduction}
The right to explanations mandates that any black-box machine learning model, making crucial decisions in high-stakes applications should provide the user, i.e., the applicant, with a suitable explanation for its decision~\cite{voigt2017eu}. In this regard, counterfactual explanations have grown as an effective means to deliver the minimum perturbation required at the user's end to alter the model's decision to a favorable outcome \cite{Harvard_discussion_first_counterfactual}. For instance, in the case of a bank loan rejection, a user might receive a counterfactual recommendation to increase their income by $10$K to get accepted. Numerous works have focused on generating counterfactuals with different properties, namely, proximity to the user's input \cite{nice}, robustness to model changes \cite{upadhyayROAR, hammanRobustCF, dutta2022robust, hammanRobustCFJournal}, feasibility under user's constraints \cite{face, dice}, sparsity in the attributes to be changed \cite{nice,dice}, and diversity of the counterfactuals \cite{dice}. We refer the reader to \cite{verma2020counterfactual, guidottiCounterfactualSurvey, mishra2021survey} for a comprehensive survey on different methods. In this work, we introduce a framework that provides information-theoretic guarantees.

Providing an appropriate counterfactual poses serious privacy concerns, both for the user asking for a counterfactual, and for the institution delivering it. Existing works such as  \cite{pawelczykMembershipInference, yang2022differentially, privacy_issue_in_cf, pentyala2023privacy} focus on preserving data privacy from the institution's side, while \cite{aivodjiModelExtraction, wangDualCFModelExtraction, dissanayakeModelExtraction} focus on the extraction of the model by querying for multiple counterfactual explanations. However, preserving privacy from the user's side has rarely received attention. We are particularly interested in the scenario where a user would like to obtain a counterfactual explanation without revealing their input feature vector to the institution. The user may be reluctant to share their feature vector with the institution for several reasons, e.g., if they have a limited number of attempts to apply, or if they wish to preserve their data privacy until they improve their chances of acceptance. 

\paragraph{Related Works:}
Since the initial formulation in \cite{Harvard_discussion_first_counterfactual}, numerous works have been focusing on generating counterfactual explanations with different properties. Proximity to the query instance \cite{nice}, robustness \cite{upadhyayROAR, hammanRobustCF}, actionability \cite{face, dice}, sparsity in change \cite{nice, dice}, and diversity \cite{dice} are some such properties; we refer the reader to \cite{karimiCFMethodsSurvey} and \cite{guidottiCounterfactualSurvey} for a comprehensive survey on different methods. In this work, we focus on both proximity to the original instance and actionability. Further, we introduce immutability as a constraint where the user may keep some of their feature values fixed, while obtaining a counterfactual. Moreover, we use nearest-neighbor counterfactuals as the explanation method considered. Existing works on privacy within the context of counterfactual explanations mainly focus on the institution's end. In this regard, \cite{pawelczykMembershipInference} analyzes inferring the membership of an explanation in the training set of the model while \cite{yang2022differentially} provides differentially private counterfactuals. \cite{aivodjiModelExtraction} and \cite{wangDualCFModelExtraction} present two ways of utilizing counterfactuals to extract the model when counterfactuals are provided for any query. \cite{dissanayakeModelExtraction} presents a model extraction strategy that specifically leverages the fact that the counterfactuals are close to the decision boundary. Explanation linkage attacks that try to extract private attributes of a nearest-neighbor counterfactual explanation are discussed in \cite{privacy_issue_in_cf}. All of these works focus on the privacy of either the model or the data stored in the institution's database. In contrast, we are interested in the privacy of the applicant who is asking for an explanation. 

Closely related to our problem is the nearest neighbor search problem, where the user needs to retrieve the indices of the vectors in a database, that are closest to their vector according to some similarity metric. In this regard, \cite{ANN_sublinear} proposes algorithms that guarantee computational privacy, both to the user and the database, while the user retrieves an approximate nearest neighbor. Reference \cite{sajani_ANN} proposes an information-theoretically private clustering-based solution based on the dot-product metric. This work, however, does not consider database privacy, and the user retrieves only an approximate nearest neighbor. The work leverages techniques from private information retrieval (PIR) \cite{chor}, which is a subject of independent interest, to obtain the index of the nearest counterfactual. 

In PIR, a user wishes to retrieve a message out of $K$ replicated messages in $N$ servers without leaking any information about their required message index. The capacity of PIR, i.e., the maximum ratio between the number of the required message symbols and the number of total downloaded symbols, is found in \cite{c_pir}. In symmetric PIR (SPIR), an extra requirement is considered where the user cannot get any information about the other messages aside from their required message. The capacity for such a model is found in \cite{c_spir}. Other important variants can be found in \cite{arbitrarycollusion,  banawan_eaves, banawan_multimessage_pir,   banawan_pir_mdscoded,  Salim_CodedPIR, batuhan_hetero,   byzantine_tpir,  grpahbased_pir, shreya_PIR,  C_SETPIR, ChaoTian,  codedstorage_adversary_tpir,  colluding, csa, first_xsecure, nomeirasymmetric, wang_spir,  uncoded_constrainedstorage_pir,  tspir_mdscoded,   utah_hetero,   tpir_sideinfo, sun_eaves,   semantic_pir, nan_eaves}. Reference \cite{csa}, proposes a cross subspace alignment (CSA) approach as a unifying framework for PIR and SPIR with additional requirements, such as security against the storing databases. These schemes are capacity-achieving in some cases, for instance, when the number of messages is large. We refer the reader to \cite{ulukusPIRLC} for a comprehensive survey on the PIR and SPIR literature. 

\paragraph{Our Contributions:} 
We summarize our contributions as follows.
\begin{itemize}
    \item First, we introduce the novel problem of private counterfactual retrieval (PCR), along with a baseline scheme to achieve user privacy to retrieve the index of the closest counterfactual using the $\ell_2$ distance metric (Theorem~\ref{PCR_baseline_thm}). Then, we develop two different PCR schemes that we call Diff-PCR and Mask-PCR to provide the institution with better privacy for their database compared to the baseline scheme while achieving perfect information-theoretic privacy for the applicant (Theorem~\ref{PCR_diff_thm} and Theorem \ref{PCR_mask_thm}).
    \item Next, we extend our PCR formulation to a more practical setting, where we allow the user to fix a subset of their features as \emph{immutable}, i.e., their values should not be altered in their corresponding counterfactual. The immutable set is user-specific and should be kept private from the institution. Towards this, we propose two immutable PCR (I-PCR) schemes (Theorem~\ref{IPCR_2phase_thm} and Theorem~\ref{IPCR_1phase_thm}), where the user retrieves their counterfactual index from the institution under the immutability constraint.
    \item Further, we account for \emph{user actionability} which guarantees preferential change of certain attributes compared to the rest. For both PCR and I-PCR, we demonstrate that our schemes can be modified to include private user actionability on the mutable attributes without compromising the privacy of the immutable attributes. We call these extended schemes PCR+ (Theorem~\ref{pcr_weighted_thm}) and I-PCR+ (Theorem~\ref{ipcr_weighted_thm}).
    \item We numerically evaluate the database leakage for the proposed PCR and I-PCR schemes. With this, we verify the comparative leakage values among the proposed schemes.
\end{itemize}   

Our framework of privately retrieving the index of the closest neighbor guarantees that the institution does not learn any information about the user’s input. Further, the institution does not learn the counterfactual explanation corresponding to the user either. This is achieved in two steps: First, we perform the PCR scheme to retrieve the counterfactual index from a set of non-colluding servers, which store the set of accepted samples through replication \cite{cassandra20192, Apache_replication}. Next, we use this index to perform symmetric private information retrieval (SPIR) \cite{c_spir} on the same indexed dataset. In SPIR, the user obtains their counterfactual explanation, without revealing its index to the server, while the server incurs no further leakage on the stored data other than the required record. The SPIR scheme applies seamlessly in the existing system, due to the replicated and non-colluding databases.

The rest of the paper is organized as follows. Section \ref{sec:prob_form} formally introduces the problem. Section \ref{sec:pcr_schemes} and Section \ref{sec:ipcr_schemes} describe the proposed schemes for the PCR and I-PCR settings, respectively. In Section \ref{sec:user_act}, we extend our schemes to include private user actionability. In Section \ref{sec:expts}, we present our numerical results. Finally, Section \ref{sec:conclude} concludes our paper.

\paragraph{Notations:} Throughout the paper, we use $[n]$ to denote the set $\{1,2,\ldots,n\}$. For all integers, $m<n$ we use $[m:n]$ to denote the set $\{m,m+1,\ldots,n-1,n\}$. Given a set $A$, we represent its cardinality by $|A|$. For a vector $a$, we write $a^T$ to denote its transpose. We use $\mathbb{F}_q$ to represent a finite field, where $q$ is a prime or a prime power. The functions $H(\cdot)$ and $I(\cdot;\cdot)$ represent the Shannon entropy and mutual information, respectively. Further, we use bold, italicized font $\bm{A}$ to denote a matrix.

\section{Problem Formulation}\label{sec:prob_form}
The institution has a pre-trained binary classification model that takes a $d$-dimensional feature vector  as input and classifies it into its target class, e.g., accepted or rejected. A user who is rejected by this model wishes to privately retrieve a valid counterfactual sample corresponding to their data sample. The user, who does not have access to the model, relies on a database $\mathcal{D}$ containing the feature vectors of a set of samples accepted by the model. 

We assume that each attribute (feature) of the samples (feature vectors) is an integer in $[0:R]$. The samples in $\mathcal{D}$ are stored in $N$ non-colluding and non-communicating servers in a replicated manner, and are indexed as $y_1, y_2, \ldots, y_M$ where $M=|\mathcal{D}|$. Under this setup, the user retrieves the index of the accepted data sample in the database closest to their own feature vector in terms of a distance metric $d(\cdot, \cdot)$. In all sections, except Section \ref{sec:user_act}, we assume the metric to be the $\ell_2$ norm, i.e., $d(x,y)=||y-x||^2$. We refer to the problem without and with the immutability constraint as the private counterfactual retrieval (PCR) setting, and private counterfactual retrieval with immutable features (I-PCR) setting, respectively, which we formally describe next.

\begin{figure}[t]
    \centering
    \includegraphics[width=0.5\textwidth]{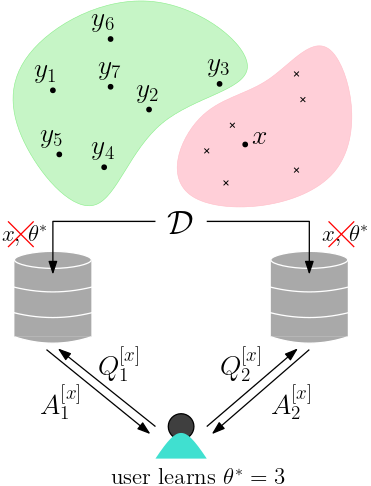}
    \caption{PCR system model: The green and red regions represent the accepted and rejected samples, respectively. The user learns that $\theta^*=3$ is their counterfactual index in $\mathcal{D}$.}
    \label{system_model_case1}
\end{figure}

\paragraph{PCR setting:} Here, we assume that the user can change all their feature values to obtain their counterfactual, and the counterfactual index $\theta^*$ is given by
\begin{align}\label{goal_pcr}
    \theta^*=\arg\min_{i\in [M]} d(x,y_i),
\end{align}
as depicted in Fig.~\ref{system_model_case1}. 

\begin{figure}[t]
    \centering
    \includegraphics[width=0.6\textwidth]{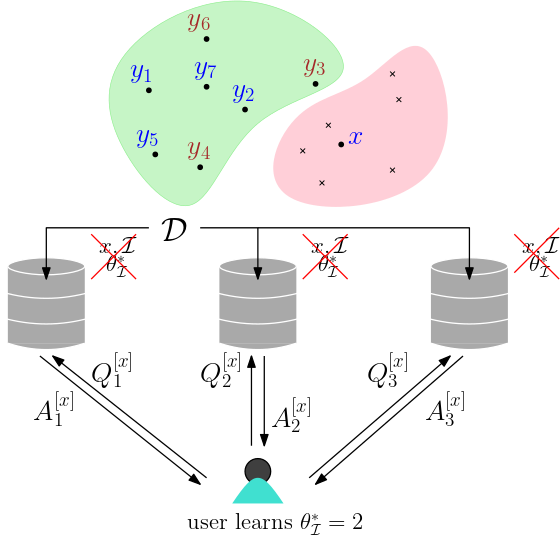}
    \caption{I-PCR system model: Here,  $y_1$, $y_2$, $y_5$, $y_7$ have the same values for the immutable features as $x$. The user learns that $\theta^*_{\mathcal{I}}=2$ is their counterfactual index in $\mathcal{D}$.}
    \label{sysmod_immutable}
    \vspace*{-0.3cm}
\end{figure}

To this end, the user sends the query $Q_n^{[x]}$ to server $n\in [N]$. Upon receiving the queries, each server computes their answer $A_n^{[x]}$ using their storage, their queries and shared common randomness $Z'$, i.e.,
\begin{align}\label{ans_gen_pcr}
    H(A_n^{[x]}|\mathcal{D},Q_n^{[x]},Z') = 0.
\end{align}
Using the responses from all the servers, the user determines the index $\theta^*$ of their corresponding counterfactual, i.e., 
\begin{align}\label{eq:decodability_pcr}
    [\text{PCR Decodability}] \quad H(\theta^*| Q_{[N]}^{[x]}, A_{[N]}^{[x]}, x) = 0,
\end{align}
where $Q_{[N]}^{[x]}$ and $A_{[N]}^{[x]}$ denote all the sent queries, and all the received answers, respectively. Since user privacy is our primary focus, we impose that each server must not acquire any information on the user's sample or the index of their counterfactual, i.e., $\forall n \in [N]$,
\begin{align}\label{eq:user_priv_pcr}
    [\text{PCR User Privacy}] \quad I(x, \theta^*; Q_n^{[x]}| \mathcal{D})=0,
\end{align}
where $I(\cdot;\cdot|\cdot)$ is the conditional mutual information. To quantify privacy for the servers, we consider a mutual information based metric that defines the amount of information leakage about the samples in $\mathcal{D}$ to the user, upon receiving the answers, as follows
\begin{align}\label{eq:leakage_pcr}
[\text{PCR Database Leakage}] \quad I(y_1,\ldots,y_M; Q_{[N]}^{[x]}, A_{[N]}^{[x]}|x).
\end{align}
\paragraph{I-PCR setting:}
In PCR with the immutability constraint, the goal of the user is to retrieve the index of the nearest sample which should differ from $x$ in at most a subset $\mathcal{M}\in [d]$ of features. Alternatively, the user fixes their counterfactual's feature values in $\mathcal{I} = [d]\setminus \mathcal{M}$, which we call the \emph{immutable set}. In particular, let $\mathcal{I}=\{j_1, j_2, \ldots, j_{|\mathcal{I}|}\}$ with $j_1<j_2<\ldots<j_{|\mathcal{I}|}$, then 
\begin{align}\label{goal_ipcr}
    x_{\mathcal{I}} = 
    \begin{bmatrix}
        x_{j_1} & x_{j_2} & \ldots & x_{j_{|\mathcal{I}|}}
    \end{bmatrix}^T,
\end{align} 
and $x_{\mathcal{M}}$ is defined in a similar way. Therefore, the goal of the user is to find the counterfactual index $\theta^*_{\mathcal{I}}$ given by
\begin{align}
    \theta^*_{\mathcal{I}}=\arg \min_{i\in \Theta} d(x,y_i)
    \quad \text{with} \quad \Theta = \{i\in [M]: y_{i,\mathcal{I}}=x_{\mathcal{I}}\},
\end{align}
where the knowledge of $\mathcal{I}$, hence that of $\Theta$, is restricted to the user and no information on $x$ is revealed to the server. The system model is illustrated in Fig.~\ref{sysmod_immutable}. Since it is clear from the context, we use the same notations $Q_n^{[x]}$ and $A_n^{[x]}$ to denote the query and answer corresponding to the $n$th server for the I-PCR setting. The answers are deterministic functions of the database, queries and shared randomness as given in \eqref{ans_gen_pcr}. The queries and answers need to satisfy the decodability, user privacy and database leakage constraints, as follows
\begin{align}\label{eq:ipcr_decodability}
    \text{[I-PCR Decodability]} & & H(\theta^*_{\mathcal{I}}| Q_{[N]}^{[x]}, A_{[N]}^{[x]}, x,\mathcal{I}) = 0,\\
    \label{eq:ipcr_privacy}
    \text{[I-PCR User Privacy]} & & I(x, \theta^*_{\mathcal{I}}, \mathcal{I}; Q_n^{[x]}, A_n^{[x]}| \mathcal{D})=0,\\
    \label{eq:ipcr_db_leakage}
    \text{[I-PCR Database Leakage]} & & I(y_{[M]}; Q_{[N]}^{[x]},A_{[N]}^{[x]}|x, \mathcal{I}).
\end{align}

Our goal is to develop schemes that reduce this database leakage, while maintaining perfect user privacy for both PCR and I-PCR settings. The performances of our proposed schemes, which will be analyzed in detail subsequently, are listed in Table~\ref{Table:performance comparison_all schemes} for convenience.

\vspace*{0.4cm}

\begin{table}[h]
    \centering
    \begin{tabular}{|c|c|c|}
    \hline
    Scheme & Communication cost & $\mathbb{F}_q$ size requirement\\
    \hline
    
    Baseline PCR ($N=2$) & $2d+2M$ & $q>R^2d$ \\
    \hline
    Diff-PCR ($N=2$) & $2d+2M-2$ & $q>2R^2d$  \\
    \hline
    Mask-PCR ($N=2$) & $2d+2M$ & $q>R^2d$ \\
    \hline
    \hline
    Two-Phase I-PCR ($N=3$) & $9d+9M$ & $q>R^2d$\\
    \hline
    Single-Phase I-PCR ($N=3$) & $6d+3M$ & $q>F(L-1)R^2+R^2d: L>R^2d$ \\
    \hline
    \hline
    Baseline PCR+  ($N=3$)& $6d+3M$ & $q>R^2L_1d$ \\
    \hline
    Diff-PCR+ ($N=3$) & $6d+3M-3$ & $q>2R^2L_1d$ \\
    \hline
    Mask-PCR+ ($N=3$) & $6d+3M$ & $q>R^2L_1d$ \\
    \hline
    \hline
    Two-Phase I-PCR+ ($N=4$) &$14d+11M$ & $q>R^2L_1d$\\
    \hline
    Single-Phase I-PCR+ ($N=3$) & $6d+3M$ & $q>F(L-1)R^2+R^2L_1d:L>R^2L_1d$\\
    \hline
    \end{tabular}
    \vspace*{0.1cm}
    \caption{Summary of the performances of the proposed scheme. We show communication cost, minimum required field size and number of servers required, to realize our proposed schemes. $R$ is the maximum value that each feature is quantized to, $d$ is the dimension of the feature vector, $M$ is the database size, $F$ is the maximum number of immutable features, and $L_1$ is the largest integer of the weight vector for actionability.}
    \label{Table:performance comparison_all schemes}
\end{table}

Before describing the schemes, we state the definition of a Vandermonde matrix, which is central to all our proposed schemes. 

\begin{definition}[Vandermonde matrix]
    Let $\alpha_1,\ldots,\alpha_n$ be distinct elements of $\mathbb{F}_q\setminus \{0\}$. Then, a Vandermonde matrix of order $n$ on $\mathbb{F}_q$ is denoted by $\bm{V}_n$, whose $(i,j)$th entry is given by
    \begin{align}
        \bm{V}_n(i,j) = \alpha_i^{j-1}, \qquad  j=[n], \quad i=[n].
    \end{align}
    Since $\alpha_1,\ldots,\alpha_n$ are non-zero and distinct, $\bm{V}_n$ is invertible in $\mathbb{F}_q$.
\end{definition}

In addition, for completeness, Shannon's one-time pad theorem, which we use extensively, is stated below as in \cite{coverthomas}.

\begin{theorem}[One-time pad]
    For any random variable $X \in \mathbb{F}_q$, and any uniform random variable $Z$ on $\mathbb{F}_q$, the following holds
    \begin{align}
        I(X; X+Z)=0,
    \end{align}
    where addition is according to $\mathbb{F}_q$ arithmetic.
\end{theorem}

\section{PCR Schemes}\label{sec:pcr_schemes}
In this section, we propose schemes that provide feasible solutions to minimizing \eqref{eq:leakage_pcr} subject to \eqref{eq:decodability_pcr} and \eqref{eq:user_priv_pcr}.
For ease of exposition, we write
\begin{align}
    d_i(x) =  ||y_i-x||^2 , \quad \forall i\in [M].
\end{align}
First, we present the baseline scheme, followed by two approaches, difference of distances (Diff-PCR) and masking (Mask-PCR) that leak less information on $\mathcal{D}$, compared to the baseline, in addition to achieving user privacy. All PCR schemes require $N=2$ replicated servers.

\subsection{Baseline PCR}\label{userpriv1}

\begin{theorem}\label{PCR_baseline_thm}
    There exists a scheme that can retrieve the exact closest counterfactual index from $N=2$ servers with perfect user privacy, with the communication cost of $2(d+M)$ symbols of $\mathbb{F}_q$ where $q>R^2d$ is prime.
\end{theorem}
  
To prove the achievability of Theorem~\ref{PCR_baseline_thm}, we present our baseline PCR scheme. Let the operating field be $\mathbb{F}_q$ where $q$ is a prime satisfying $q>R^2d$. Each server stores the $M$ accepted samples, $y_1, \ldots, y_M$, each being a $d$ dimensional vector. Let $\alpha_1, \alpha_2$ be two distinct elements of $\mathbb{F}_q$ globally known to the user and the servers. The user privately generates a random vector $Z$ uniformly from $\mathbb{F}_q^d$ and then sends the query $Q_n^{[x]}$ to the $n$th server based on $x$ as
\begin{align}\label{queries_no_weight}
    Q_n^{[x]} = x+\alpha_n Z, \quad n=1,2.
\end{align}
Since $x$ is one-time padded with $\alpha_nZ$ which is uniform on $\mathbb{F}_q^d$, each server learns no information about $x$, i.e., $I(x;Q_n^{[x]}|y_{[M]}) = 0$. Let the servers share $M$ independent random variables $Z'(i), i \in [M]$ picked uniformly from $\mathbb{F}_q$. Given a query, the servers compute one answer for each $y_i\in \mathcal{D}$ as follows
\begin{align}
    A_n^{[x]} (i) =& ||y_i-Q_n^{[x]}||^2 + \alpha_n Z'(i)  \\
    =& d_i(x)+  \alpha_n \left(-2(y_i-x)^T Z + Z'(i)\right) + \alpha_n^2 ||Z||^2.
\end{align}
Note that $A_n^{[x]} (i)$ is a second degree polynomial of $\alpha_n$ with $\alpha_n^2 ||Z||^2$ already known to the user. Therefore, the user cancels $\alpha_1^2||Z||^2$ and $\alpha_2^2||Z||^2$ from $A_1^{[x]}(i)$ and $A_2^{[x]}(i)$, respectively, to obtain
\begin{align}
    \label{eq: ans_reprocess}
    \begin{bmatrix}
        \hat{A}_1^{[x]}(i)\\
        \hat{A}_2^{[x]}(i)
    \end{bmatrix}
    &=\begin{bmatrix}
        A_1^{[x]}(i)-\alpha_1^2||Z||^2\\
        A_2^{[x]}(i)-\alpha_2^2||Z||^2
    \end{bmatrix}\\ 
    &=\bm{V}_2
    \begin{bmatrix}
        d_i(x)\\
        -2 (y_i-x)^T Z + Z'(i)
    \end{bmatrix}, \label{get_d_i(x)}
\end{align}
where $\bm{V}_2$ is the two-dimensional Vandermonde matrix given in this case as $\bm{V}_2=\begin{bmatrix} 1 & \alpha_1 \\ 1 & \alpha_2 \end{bmatrix}$. 

After multiplying (\ref{get_d_i(x)}) with $\bm{V}_2^{-1}$, the user obtains $d_i(x)$ for all $i\in[M]$. The user then compares the values of $d_i(x)$, for all $i\in [M]$ and assigns $\theta^*$ to the $i$ for which this is minimum. 

\paragraph{Communication Cost:} This scheme requires two $d$-dimensional vectors of $\mathbb{F}_q$ to be sent, one to each server. This entails an upload cost of $2d$ symbols. Each server responds with $M$ symbols from $\mathbb{F}_q$, thereby, incurring a download cost of $2M$.

\paragraph{Computation Complexity:} {At the user side, the computation for the queries involves two scalar multiplications and additions in $d$ dimensions, which is $O(d)$. To find $\theta^*$, the user decodes $d_i(x)$ using \eqref{get_d_i(x)} for each $i\in [M]$ and compares them to find the minimum, both using $O(M)$ operations. Therefore, the user-side computation complexity is $O(d+M)$. Each server generates $M$ answers, where each answer involves subtraction, squaring, and addition of $d$ elements of each vector. This results in $O(Md)$ operations at each server.}

\begin{remark}
    Note that the distance metric in the system model and all proposed schemes use the $\ell_2$ norm. The system model and the schemes can be extended to any $\ell_k$ norm, where $k$ is even by modifying the answer generation accordingly. We demonstrate this for the baseline PCR scheme in Appendix ~\ref{apend_high_norm}. 
\end{remark}

\subsection{Diff-PCR}

\begin{theorem}\label{PCR_diff_thm}
    There exists a scheme  with a lower value of database leakage \eqref{eq:leakage_pcr} compared to the baseline PCR, with the communication cost of $2(d+M-1)$ symbols of $\mathbb{F}_q$ where $q>2R^2d$ is prime.
\end{theorem}

The main idea of the scheme is to improve database privacy by revealing only the difference of the norms, as opposed to all the norms, while maintaining the user's privacy. We show that this can be accomplished with $N=2$ replicated databases as in the baseline PCR. The field of operation is a prime $q> 2R^2d$. This is required because for all $i,j\in [M]$,
\begin{align}
    0\leq| d_i(x)- d_j(x)|\leq R^2d.
\end{align}
Therefore, 
\begin{align}
    d_i(x)- d_j(x)\in
    \begin{cases} 
        [0:R^2d], & d_i(x)\geq d_j(x),\\
        [R^2d+1:q-1], &  d_i(x)<d_j(x).
    \end{cases}
\end{align}
The servers share a common random vector $Z'_{[M-1]}=[Z'(1) , Z'(2), \ldots, Z'(M-1)]^T$, where each entry is picked uniformly and independently from $\mathbb{F}_q$. As described in Section \ref{userpriv1}, the user sends the query given in \eqref{queries_no_weight} to server $n=1,2$. Then, server $n$ computes the following answer for each $i\in [M-1]$,
\begin{align}
    A_n^{[x]}(i)=&||y_i-Q_n^{[x]}||^2 - ||y_{i+1}-Q_n^{[x]}||^2 + \alpha_n Z'(i)\\
    =&||y_i-x||^2 - ||y_{i+1}-x||^2 +\alpha_n\left(2(y_{i+1}-x)^T Z-2(y_i-x)^T Z+ Z'(i)\right) \\
    =& d_i(x)- d_{i+1}(x) +\alpha_n I'(i),
\end{align}
where $I'(i)=2(y_{i+1}-y_i)^T Z+ Z'(i)$. Using $A^{[x]}_1(i)$ and $A^{[x]}_2(i)$, the user exactly recovers $||y_i-x||^2 - ||y_{i+1}-x||^2$, because
\begin{align}\label{get_diff_d_i}
    \begin{bmatrix}
        A_1^{[x]}(i)\\
        A_2^{[x]}(i)
    \end{bmatrix}=
    \bm{V}_2
    \begin{bmatrix}
        d_i(x)-d_{i+1}(x)\\
        I'(i)
    \end{bmatrix}.
\end{align}
From here, the user recovers the $M-1$ differences along with $M-1$ interference terms. In each $I'(i)$, the one-time-padding with $Z'(i)$ makes sure that no information on $y_{i+1}-y_i$ is revealed. The user finds the index $\theta^*$ of their counterfactual using Algorithm \ref{alg_difference_distance}.

\begin{algorithm}
    \caption{Algorithm to compute $\theta^*$}
    \label{alg_difference_distance}
    \hspace*{\algorithmicindent} \textbf{Input:} $d_i(x)-d_{i+1}(x), \quad i \in [M-1]$\\
    \hspace*{\algorithmicindent} \textbf{Output:} $\theta^*$
    \begin{algorithmic}[1]
        \State {$\theta^* = 1$}
        \For {$i \in [M-1]$}
        \State {$r(i)= d_i(x) - d_{i+1}(x)$}
        \State {$d_{\theta^*}(x) - d_{i+1}(x)= \sum_{j=\theta^*}^{i}r(j)$}
        \If {$d_{\theta^*}(x) - d_{i+1}(x)\in [R^2d+1:q-1]$}\label{range_prox}
        \State {$\theta^* \leftarrow \theta^*$}
        \Else
        \State {$\theta^* \leftarrow i+1$}
        \EndIf
        \EndFor
        \State \Return $\theta^*$
    \end{algorithmic}
\end{algorithm}

\paragraph{Communication Cost:} The upload cost incurred in this scheme is $2d$ and the download cost is $2(M-1)$.

\paragraph{Computation Complexity:} The user side query generation is $O(d)$ as in the baseline scheme. To find $\theta^*$, the user finds $d_i(x)-d_{i+1}(x)$ using \eqref{get_diff_d_i} and evaluates $\theta^*$ using Algorithm \ref{alg_difference_distance}, which has a complexity of $O(M)$. Thus, the user's computation complexity is $O(d+M)$. The server side complexity is $O(Md)$, same as the baseline.

\subsection{Mask-PCR}\label{masking_equal_actionable}

\begin{theorem}\label{PCR_mask_thm}
    There exists a scheme that has a lower database leakage in terms of \eqref{eq:leakage_pcr} compared to the baseline PCR with the communication cost of $2(d+M)$ symbols of $\mathbb{F}_q$ where $q>R^2d$ is prime.
\end{theorem}

In this approach, we require the servers to have access to the set of rejected samples, $\mathcal{D}_{r} = \{x_1, \ldots, x_K\}$ (this restriction is removed during the experimental analysis). Now, define the closure of $\mathcal{D}_{r}$ as follows
\begin{align}\label{remark_edit_2}
    \mathcal{D}_c= \mathtt{clo}(\mathcal{D}_r) = \{&x \in \mathbb{F}_q^d: | d_{i}(x) - d_{j}(x) |-|d_{i}(x_k) - d_{j}(x_k)| \geq 0, \forall i, j, k\}\setminus \mathcal{D}.
\end{align}
In addition, we define the following metrics 
\begin{align}\label{remark_edit}
    d_{k} = \min_{i,j\in [M]} |d_{i}(x_k) - d_{j}(x_k)|, \qquad d_{\min} = \min_{k\in [K]} d_k.
\end{align}
Let $\mu$ be a random variable with support $\{0,\ldots, d_{\min}-1\}$. Now, a user who wishes to know the closest accepted sample to their rejected sample $x \in \mathcal{D}_c$ sends the query in \eqref{queries_no_weight} to the $n$th server. Upon receiving the query, each server computes its answer for each $i \in [M]$ as follows
\begin{align}
    A_n^{[x]}(i) = ||y_{i}-Q_n^{[x]}||^2 + \mu(i) + \alpha_n Z'(i), 
\end{align}
where $\mu(i)$ has the same distribution as $\mu$ and $Z'(i)$ is a uniform random variable in $\mathbb{F}_q$. As the user receives the answers from the servers, they are reprocessed according to \eqref{eq: ans_reprocess} to decode the distances as follows
\begin{align}
    \begin{bmatrix}
        \hat{A}_1^{[x]}(i)\\
        \hat{A}_2^{[x]}(i)
    \end{bmatrix}= \bm{V}_2\begin{bmatrix}
        d_i(x) + \mu(i)\\
        I(i)+ Z'(i)
    \end{bmatrix}.
\end{align}
Upon getting the values of the masked distances, $d_i(x) + \mu(i)$, $i \in [M]$, the user utilizes them to decide which is closest based on their numerical value, i.e., decode the index of the closest accepted sample. To show that the user can correctly decode with the masked distances, we need the following lemma.

\begin{lemma}\label{Mask-PCR-decoding-lemma}
    If $x \in \mathcal{D}_c$, then the user is able to decode the index of the closest acceptable sample.
\end{lemma}

Here, in contrast to the Diff-PCR approach, the field size does not need to be expanded. The main reason for this is that the difference calculations in this case are done at the user side and therefore the field size restriction can be dropped. The proof of Lemma~\ref{Mask-PCR-decoding-lemma} is provided in Appendix~\ref{lemma_mask_pcr_proof}.

Lemma~\ref{Mask-PCR-decoding-lemma} states that all the vectors in $\mathcal{D}_c$ preserve the relative distances among the accepted samples, i.e., the ordering of the distances among the points in $\mathcal{D}_c$ and accepted samples is maintained. In addition, note that the user does not know the exact value of $d_{\min}$ since they do not have any prior knowledge of the accepted samples.    

\paragraph{Communication Cost:} The upload cost is $2d$ and the download cost is $2M$.

\paragraph{Computation Complexity:} The computation complexity at the user and the server is equal to that of the baseline PCR, i.e., $O(d+M)$ and $O(Md)$, respectively.

\begin{remark}
    It is important to note that \eqref{remark_edit} is evaluated prior to the scheme initiation and done once only till the dataset itself changes. This is a part of pre-processing for Mask-PCR and is not a part of the main scheme, which begins with the user sending their query. This is why it is not included in the calculation of the computation complexity. 
\end{remark}

\begin{remark}\label{rmk:closure_lift}
    We believe that the requirement $x\in \mathcal{D}_c$ is a bit strong, which is why we dropped this requirement in our first experiment (see Sec \ref{exp:accu_quant}) and we find $d_{\min}$ empirically instead. However, it is required for theoretical analysis to guarantee decodability.
\end{remark}

\paragraph{Effect of Field Size in Masking:}
The field size can have a significant role in Mask-PCR. Recall that the minimum requirement for the field size is $R^2d$ as explained at the beginning of this section. Let $q_1$ be the field size satisfying $q_1> R^2 d$. If another field size $q_2$ is chosen such that $q_2 > q_1$, we can embed the samples $y_1, \ldots, y_M$ in $\mathbb{F}_{q_2}$ using a relative distance preserving transform $T: \mathbb{F}_{q_1} \rightarrow \mathbb{F}_{q_2}$ such that the following condition is satisfied
\begin{align}
    |d_{i}(x_k)-d_{j}(x_k)| \leq | \big(||x_k -T(y_i)||^2- ||x_k -T(y_j)||^2\big)|,
    \qquad k\in [K], \quad i,j\in [M].
\end{align}
The support of the random variable used in masking $\mu$ is larger compared to when the field size is $q_1$ since $d_{\min} \leq |d_{i}(x_k)-d_{j}(x_k)|$. Thus, the estimation error of the user for the exact values of the samples can increase. In addition, note that this transform can be kept hidden from the user since it does not affect the result for ordering of the samples by construction.

\paragraph{Illustrative Example:}
This example demonstrates the masking approach and the field size expansion. Let the rejected samples, for example, be $\{[1,2]^T, [2,1]^T\}$, and the set of accepted samples be $\{[20,0]^T,[0,20]^T\}$. Let the field size be $q_1 = 809$. Thus, after simple calculations, we see that the range of the masking random variable is $\{0,\ldots,39\}$. Let the user choose $x=[1,2]^T$. Thus, the queries sent to the two servers are given by \eqref{queries_no_weight}
and after the servers reply, the user reprocesses the received answers according to \eqref{eq: ans_reprocess}, to obtain
\begin{align}
   \begin{bmatrix}
        \hat{A}_1^{[x]}(1)\\
        \hat{A}_2^{[x]}(1)
    \end{bmatrix} = \bm{V}_2\begin{bmatrix}
        365+\mu(1)\\
        I(1)+Z'(1)
    \end{bmatrix},
\end{align}
and
\begin{align}
  \begin{bmatrix}
        \hat{A}_1^{[x]}(2)\\
        \hat{A}_2^{[x]}(2)
    \end{bmatrix} = \bm{V}_2\begin{bmatrix}
        325+\mu(2)\\
        I(2)+Z'(2)
    \end{bmatrix}.
\end{align}
It is clear that for any choice of $\mu(1)$, and $\mu(2)$, the user would be able to know that $y_2$ is closest to $x$.

Now, let us choose a larger field size, for example, $q_2 = 40009$ and apply the transform $T:a\rightarrow 10a$. Then, with simple calculations, we see that the range of the masking random variable is $\{0,\ldots,399\}$ which is larger than the previous range and still maintains the relative distance. We encourage the reader to refer to Section \ref{exp:accu_quant} for the effect of field size and $d_{\min}$ on the accuracy of the derived counterfactual.

\subsection{Database Leakage}
Now, we analyze the database leakage of the Diff-PCR and Mask-PCR schemes, and show that both schemes incur lower leakage than the baseline PCR scheme. The database leakage in \eqref{eq:leakage_pcr} can be rewritten as follows
\begin{align}
    I(y_{[M]}; Q_{[N]}^{[x]}, A_{[N]}^{[x]}|x) 
    &=I(y_{[M]}; A_{[N]}^{[x]}|x, Q_{[N]}^{[x]})+ I(y_{[M]}; Q_{[N]}^{[x]}|x) \\
    &=I(y_{[M]}; A_{[N]}^{[x]}|x, Q_{[N]}^{[x]}) \label{eq:queries_db_indep}\\
    &= I(y_{[M]}; A_{[N]}^{[x]}|x). \label{eq_dropping_Q}
\end{align}
where \eqref{eq:queries_db_indep} follows since $y_{[M]}=(y_1,\ldots,y_M)$ are independent of the queries $Q_{[N]}^{[x]}$ given $x$. Also, we can remove  
$Q_{[N]}^{[x]}$ from the conditioning terms as in \eqref{eq_dropping_Q}, simplifying the leakage expression. This is because, the randomness symbols in $Q_{[N]}^{[x]}$ added for the privacy of $x$, upon answer generation, are protected by the randomness shared among the servers. Additionally, they align on a subspace different from the database information subspace, leading to no leakage from the interference terms. The database leakage for the baseline scheme is therefore, 
\begin{align}
    I(d_1(x), \ldots, d_M(x) ;y_{[M]} | x).
\end{align}
Let $r(i)=d_{i+1}(x)-d_i(x)$. Then, for the Diff-PCR scheme, it holds that
\begin{align}
    I(&r(1), r(2),\ldots, r(M-1); y_{[M]} |x) 
   \leq  I(d_1(x), \ldots, d_M(x) ;y_{[M]} | x), \label{cmt_3}
\end{align}
due to the data-processing inequality. This implies that the database leakage is lower than that of the baseline scheme.

To compare the leakages of the masking and the baseline schemes, we proceed as 
\begin{align}
    I(y_{[M]} ; & d_i(x)+ \mu(i),i \in [M]| x) \nonumber\\
    =& H(d_1(x) + \mu(1), \ldots, d_M(x) + \mu(M) | x) -H(d_i(x)+\mu(i), i \in [M]|y_{[M]}, x)\\
    \leq& H(d_1(x) , \mu(1), \ldots, d_M(x) , \mu(M) | x) -H(\mu(i), i \in [M]|y_{[M]}, x)\\
    =& H(d_1(x), \ldots, d_M(x) | x)  + H(\mu(1), \ldots, \mu(M) | x,d_1(x) , \ldots, d_M(x) )\nonumber \\
    &-H(\mu(1),  \ldots, \mu(M)|y_{[M]}, x)\label{eq:mu_deterministic}\\
    =&  H(d_1(x), \ldots, d_M(x) | x)  - H(d_1(x), \ldots, d_M(x) | x, y_{[M]}))\label{eq:mu_deterministic_reduced}\\
    =& I(y_{[M]}; d_1(x), \ldots, d_M(x) | x),
\end{align}
where \eqref{eq:mu_deterministic_reduced} follows since the last two terms in  \eqref{eq:mu_deterministic} are equal, and $H(d_1(x), \ldots, d_M(x) | x, y_{[M]})$ is zero.
\begin{remark}
    Observe that the two schemes are not analytically comparable. This is because of the parameter $d_{\min}$ in the Mask-PCR scheme, which affects the amount of leakage, since the masking random variable $\mu$ depends on it. Moreover, $d_{\min}$ depends on $\mathcal{D}$, which is arbitrarily distributed. To resolve this, we numerically evaluate the leakage for different values of $d_{\min}$ and compare this with the Diff-PCR scheme (see Section \ref{expt:leakage_pcr}). 
\end{remark}

\section{I-PCR Schemes}\label{sec:ipcr_schemes}
In this section, we propose schemes that provide feasible solutions to minimizing \eqref{eq:ipcr_db_leakage} subject to \eqref{eq:ipcr_decodability} and \eqref{eq:ipcr_privacy}. Further, let $d_i(x) =  ||y_i-x||^2, i\in \Theta$.
First, we propose a scheme that operates in two phases, where the user starts by privately retrieving $\Theta$, and uses this to prepare queries for the next phase. Next, we propose a single-phase scheme, which incurs a lower communication cost than the two-phase scheme, albeit requires a larger operating field. Finally, we show that the database leakage in the two-phase scheme is lower than that in the single-phase scheme. All I-PCR schemes require $N = 3$ replicated servers.

\subsection{Two-Phase I-PCR}\label{subsec:ipcr_2phase}

\begin{theorem}\label{IPCR_2phase_thm}
    There exists a two-phase scheme that can retrieve the exact closest counterfactual index, under the immutability constraint, from $N=3$ servers with perfect user privacy, with the communication cost of $9(d+M)$ symbols of $\mathbb{F}_q$ where $q>R^2d$ is prime.
\end{theorem}

The scheme proceeds in two phases. In the first phase, the user retrieves $\Theta$, i.e., the indices $i$ of $y_i$ for which ${x}_{\mathcal{I}} = y_{i,\mathcal{I}}$. In the second phase, the user retrieves the squared $\ell_2$ distance of the samples indexed by $\Theta$ and compares them to determine the counterfactual index $\theta^*_{\mathcal
{I}}\in \Theta$. The scheme operates in a field $\mathbb{F}_q$ where $q>R^2d$ is prime. Further, the set of queries and answers for the $n$th server over two phases  are $Q_n^{[x]}=[Q_n^{[x_{\mathcal{I}}]}, Q_n^{[x_{\mathcal{M}}]}]$ and $A_n^{[x]}=[A_n^{[x_{\mathcal{I}}]}, A_n^{[x_{\mathcal{M}}]}]$, respectively.

\paragraph{Phase 1:} Based on the user's choice of $\mathcal{I}$, the user constructs the binary vector $h_1\in \{0,1\}^{d}$ such that its $k$th entry is
\begin{align}
    h_1(k) = \mathbbm{1}\left[k \in \mathcal{I}\right], \quad k\in [d].
\end{align}
Then, the user sends the following query tuple to server $n$
\begin{align}
    Q_n^{[x_{\mathcal{I}}]} = &\begin{bmatrix} Q_n^{[x_{\mathcal{I}}]}(1) & Q_n^{[x_{\mathcal{I}}]}(2)\end{bmatrix} \\ 
    =&\begin{bmatrix} h_1+\alpha_n Z_1 & x\circ h_1 + \alpha_n Z_2 \end{bmatrix},
\end{align}
where $\circ$ denotes element-wise product, and $Z_1, Z_2$ are chosen uniformly and independently from $\mathbb{F}_q^{d}$ by the user. Upon receiving the query, server $n$ responds with the following answer corresponding to each $i\in [M]$
\begin{align}
    A_n^{[x_{\mathcal{I}}]}(i) =& \rho_i|| Q_n^{[x_{\mathcal{I}}]}(1)\circ y_i - Q_n^{[x_{\mathcal{I}}]}(2)||^2 + \alpha_n Z_1'(i)+ \alpha_n^2 Z_2'(i),
\end{align}
where $\rho_i$ is chosen uniformly at random from $\mathbb{F}_q\setminus\{0\}$ by the servers, $Z_1'(i)$ and $Z_2'(i)$ are uniform random variables chosen independently from $\mathbb{F}_q$. Thus,
\begin{align}\label{answer_twophase}
     A_n^{[x_{\mathcal{I}}]}(i) =& \rho_i || (h_1+\alpha_n Z_1)\circ y_i - (x\circ h_1 + \alpha_n Z_2)||^2 + \alpha_n Z_1'(i)+ \alpha_n^2 Z_2'(i)\\
     =& \rho_i ||h_1\circ (y_i - x)||^2 + \alpha_n \big(Z'_1(i)+2 \rho_i (h_1\circ (y_i - x))^T(Z_1\circ y_i-Z_2)\big)\notag \\
     &+ \alpha_n ^2 \big( \rho_i||Z_1\circ y_i - Z_2||^2+ Z'_2(i)\big). 
\end{align}
From the answers of the three servers, the user obtains the answer vector $A^{[x_{\mathcal{I}}]}(i)$ as
\begin{align}\label{answer_phase1}
   \begin{bmatrix}
        A_1^{[x_{\mathcal{I}}]}(i)\\
        A_2^{[x_{\mathcal{I}}]}(i)\\
        A_3^{[x_{\mathcal{I}}]}(i)
    \end{bmatrix}
    =
    \bm{V}_3 
    \begin{bmatrix}
        \rho_i||h_1\circ (y_i - x)||^2 \\
        I_{11}(i)\\
        I_{12}(i)
    \end{bmatrix},
\end{align}
where $I_{11}(i)$, $I_{12}(i)$ are interference terms that reveal no information on $\mathcal{D}$ to the user. For each $i$, the user concludes that, $i \in \Theta$ iff
\begin{align}
    \rho_i||h_1\circ (y_i - x)||^2 = 0.
\end{align}

A non-zero value of  $\rho_i||h_1\circ (y_i - x)||^2$ implies that $y_{i,\mathcal{I}}\neq x_{\mathcal{I}}$ since $\rho_i$ is a non-zero element of $\mathbb{F}_q$. Further, since $\rho_i$s are independent random variables, the user does not learn which samples in $\mathcal{D}$ share some immutable attributes, unless they are identical to $x_{\mathcal{I}}$. If $\Theta=\emptyset$, then the user learns that there are no valid counterfactuals in $\mathcal{D}$ satisfying their current requirement, and the scheme ends here. Further, if $|\Theta|=1$, the user does not need the second phase.

\paragraph{Phase 2:} The user now compares the distances in a subset of $\mathcal{D}$, i.e., $\{y_i: i\in \Theta\}$ to find their counterfactual index $\theta^*$ without revealing $\Theta$ to the servers. Towards this, the user prepares the following binary vector $h_2 \in \{0,1\}^{M}$
\begin{align}\label{define_h2}
    h_2 (i) = 
    \begin{cases}
        1, & \text{ if $i \in \Theta$},\\
        0, & \text { otherwise.}
    \end{cases}
\end{align}
Then, the user sends the following query tuple to server $n$
\begin{align}
    Q_n^{[x_{\mathcal{M}}]} = & \begin{bmatrix} Q_n^{[x_{\mathcal{M}}]}(1) &  Q_n^{[x_{\mathcal{M}}]}(2)\end{bmatrix} \\ 
    = & \begin{bmatrix}h_2 + \alpha_n Z_3 & x + \alpha_n Z_4\end{bmatrix},
\end{align}
where $Z_3$ and $Z_4$ are uniform vectors from $\mathbb{F}_{q}^{M}$ and $\mathbb{F}_{q}^{d}$, respectively. Each server constructs their masked database $\tilde{S}_n$ as follows
\begin{align}
    \tilde{S}_n &= \text{diag}(Q_n^{[x_{\mathcal{M}}]}(1)) \cdot \begin{bmatrix}         y_1 & y_2 & \ldots & y_M
    \end{bmatrix}^T\\
    &=\begin{bmatrix}
        (h_2(1)+\alpha_nZ_3(1)) \cdot y_1^T\\
         (h_2(2)+\alpha_nZ_3(2))\cdot y_2^T\\
         \vdots\\
         (h_2(M)+\alpha_nZ_3(M))\cdot y_M^T
    \end{bmatrix}.\label{define_masked_db}
\end{align}

Let $\tilde{S}_n(i)$ denote the $i$th row of the masked database at server $n$. Then, their corresponding answer is given by
\begin{align}
    A_n^{[x_{\mathcal{M}}]} (i) = &||\tilde{S}_n (i)^T - Q_n^{[x_{\mathcal{M}}]}(2)||^2 + \alpha_n Z'_3(i) + \alpha_n^2 Z'_4(i)\\
    =& ||(h_2(i) + \alpha_n Z_3(i)) y_{i} - (x+ \alpha_n Z_4)||^2+ \alpha_n Z'_3(i) + \alpha_n^2 Z'_4(i)\\
    =& ||h_2(i) y_{i} - x||^2 + \alpha_n \big(2(h_2(i)y_{i} -x)^T(Z_3(i) y_{i} - Z_4) + Z'_3(i)\big) \notag \\
    &+ \alpha_n^2\big(||Z_3(i)y_{i} - Z_4||^2 + Z'_4(i)\big),
\end{align}
where $Z_3'(i)$ and $Z_4'(i)$ are uniform random variables from $\mathbb{F}_q$. From the answers of all three servers, the user computes the answer vector
\begin{align}\label{answer_phase2}
        \begin{bmatrix}
        A_1^{[x_{\mathcal{M}}]}(i)\\
        A_2^{[x_{\mathcal{M}}]}(i)\\
        A_3^{[x_{\mathcal{M}}]}(i)
    \end{bmatrix} = \bm{V}_3
    \begin{bmatrix}
        ||h_2(i) y_{i} - x||^2 \\
        I_{21}(i)\\
        I_{22}(i)
    \end{bmatrix},
\end{align}
where the first term reduces to
\begin{align}\label{phase2_ipcr_leakage}
 ||h_2(i) y_{i} - x||^2=
 \begin{cases}
     ||y_{i} - x||^2=d_i(x), & i\in \Theta,\\
     ||x||^2, & \text{otherwise.}
 \end{cases}
\end{align} 
In other words, if the values of the immutable features do not match, the user does not learn any information (not even their distances) on the samples $y_i$, $i\in [M]\setminus \Theta$. 

\paragraph{Communication Cost:} The above scheme incurs an upload cost of $9d+3M$ and a download cost of $6M$. Thus, the worst case communication cost, i.e., when $|\Theta|>1$ is $9d+9M$.

\begin{remark}
     Consider an alternative two-phase scheme. Instead of sending $Q_n^{[x_{\mathcal{I}}]}$ and $Q_n^{[x_{\mathcal{M}}]}(2)$, the user can ask the servers to compute $x\circ h_1$ by securely sending $x$ and $h_1$ in the query tuple, $Q_n^{[x_{\mathcal{I}}]} = \begin{bmatrix}h_1+\alpha_n Z_1 & x+\alpha_n Z_2\end{bmatrix}$. Server $n$ computes the product $(h_1+\alpha_n Z_1)\circ (x+\alpha_n Z_2) = x\circ h_1 + \alpha_n \hat{Z}_1 + \alpha_n^2 \hat{Z}_2$ where $\hat{Z}_1$ and $\hat{Z}_2$ are the resulting interference terms. Moreover, the first term in the answer \eqref{answer_twophase} is $\rho_i || (h_1+\alpha_n Z_1)\circ y_1 - (x\circ h_1 + \alpha_n \hat{Z}_1 + \alpha_n^2 \hat{Z}_2)||^2$ (added to $\alpha_n Z'_1(i)+ \alpha_n^2 Z'_2(i) + \alpha_n^3 Z'_3(i)$) leading to a polynomial of degree $4$ in $\alpha_n$. However, since $\alpha_n^4||\hat{Z}_2||^2, n\in[4]$ is known to the user, it can be canceled from the answers resulting in a reduced degree of $3$. This requires $N=4$ (instead of $3$) replicated servers to retrieve $\Theta$ and the communication cost in the first phase is $8d+4M$. In the second phase, the user sends a single query $Q_n^{[x_{\mathcal{M}}]}=Q_n^{[x_{\mathcal{M}}]}(1)=h_2+\alpha_n Z_3$ to any $3$ out of the $4$ servers who then compute their answers as described in the aforementioned scheme. This results in an upload cost of $3M$ and a download cost of $3M$. The total communication cost is, therefore, $8d+10M$. Therefore, although the scheme requires an additional server, its  communication cost is lower than the previous scheme,  if $d>M$.
\end{remark}

\paragraph{Computation Complexity:} At the user side, the computation of query vectors in the first phase is $O(d)$ and that in the second phase is $O(M)$. The decoding in both phases requires $O(M)$ operations. For each server, the answer generation requires $O(Md)$ computations in both phases.

\subsection{Single-Phase I-PCR}\label{ipcr_single_phase}

\begin{theorem}\label{IPCR_1phase_thm}
There exists a single-phase scheme that can retrieve the exact closest counterfactual index from $N=3$ servers with perfect user privacy, with the communication cost of $6d+3M$ symbols of $\mathbb{F}_q$ where $q>F(L-1)R^2 + R^2d$ is prime. Here, $F$ is an upper bound on $|\mathcal{I}|$ and $L$ is a globally known parameter.  
\end{theorem}

For this scheme, we assume that the user assigns at most $F$ features as immutable, where $F$ is globally known. Thus, $\mathcal{I}\subset [d]$ where $|\mathcal{I}|\leq F$. 
To this end, the servers and the user agree on $\mathbb{F}_q$ with $q>F(L-1)R^2+R^2d$ and prime. The knowledge of $F$ is necessary only to reduce the field size required for the scheme design. The user chooses an integer $L>R^2 d$ as a scaling factor for the immutable attributes, where the lower bound on $L$ is crucial for the resolvability of the sets $\Theta$ and $[M]\setminus \Theta$ at the time of decoding. To save on the communication cost, we let $L=R^2d+1$. Next, the user designs a vector $h\in \{1,L\}^d$
\begin{align}\label{weight_assignment}
    h(k) = 
    \begin{cases}
        L, & \text{ if $k\in \mathcal{I}$},\\
        1, & \text{ otherwise.}
    \end{cases}
\end{align}
To server $n$, the user sends the query tuple
\begin{align}
    Q_n^{[x]}=&\begin{bmatrix} Q_n^{[x]}(1) & Q_n^{[x]}(2) \end{bmatrix} \\ 
    =&\begin{bmatrix}x+\alpha_n Z_1 & h+\alpha_n Z_2\end{bmatrix},
\end{align}
where $Z_1$ and $Z_2$ are independent uniform random vectors in $\mathbb{F}_q^d$ that are private to the user. In response, server $n$ computes the following answer corresponding to $y_i$
\begin{align}
    A_n^{[x]}(i)=& (y_i - Q_n^{[x]}(1))^T\big((y_i - Q_n^{[x]}(1))\circ Q_n^{[x]}(2) \big)+\alpha_n Z'_1(i) + \alpha_n^2 Z'_2(i)\\
    =& (y_i-x)^T \big((y_i-x)\circ h\big) + \alpha_n I_1(i) +\alpha_n ^2 I_2(i)+ \alpha_n^3 Z_1^T(Z_1\circ Z_2),
\end{align}
where
\begin{align}
    I_1(i) =&(y_i - x)^T\big((y_i - x)\circ Z_2\big)- 2(y_i - x)^T(Z_1 \circ h\big) +Z'_1(i)\\
    I_2(i) =& Z_1^T\big(Z_1\circ h\big) - 2 Z_1^T\big((y_i - x)\circ Z_2\big)+Z_2'(i).
\end{align} Upon receiving the answers from the 3 servers, the user cancels the term of the third degree, since it is known to them, i.e.,
\begin{align}
    \hat{A}^{[x]}_n(i) = A_n^{[x]}(i) - \alpha_n^3  Z_1^T(Z_1\circ Z_2),
\end{align}
which gives
\begin{align}\label{answer_single_phase}
     \begin{bmatrix}
        \hat{A}^{[x]}_1(i)\\
        \hat{A}^{[x]}_2(i)\\
        \hat{A}^{[x]}_3(i)
    \end{bmatrix} = \bm{V}_3\begin{bmatrix}
        (y_i-x)^T \big((y_i-x)\circ h\big)\\
        I_1(i)\\
        I_2(i)
    \end{bmatrix}.
\end{align}
Now, to determine $\Theta$, the user checks the range of $(y_i-x)^T \big((y_i-x)\circ h\big)$. If it lies in $[0:(d-|\mathcal{I}|)R^2]$, all immutable features match in value and $i\in \Theta$, satisfying
\begin{align}
    (y_i-x)^T \big((y_i-x)\circ h\big) = d_i(x).
\end{align}
Among the indices in $\Theta$, the user selects the index with the minimum value of $d_i(x)$ as $\theta^*_{\mathcal{I}}$. On the other hand, if at least one immutable feature does not match, the resulting range of $(y_i-x)^T \big((y_i-x)\circ h\big)$ is $[L: L|\mathcal{I}|R^2 + (d-|\mathcal{I}|)R^2]$. Since $L$ is chosen such that $L>R^2d$, these two sets are disjoint and the user is able to distinguish between these samples.

\paragraph{Communication Cost:}  The upload cost is $2d$ symbols to each server, while the download cost is $M$ symbols from each server. This results in a communication cost of $6d+3M$.

\paragraph{Computation Complexity:} The number of computations is $O(M+d)$ at the user side, and $O(Md)$ at each server side.

\subsection{Database Leakage}
Now, we compare the database leakage incurred by both the schemes. The database leakage in \eqref{eq:ipcr_db_leakage} can be rewritten as follows
\begin{align}
 I(y_{[M]}; Q_{[N]}^{[x]}, A_{[N]}^{[x]}|x, \mathcal{I})
 &=I(y_{[M]}; A_{[N]}^{[x]}|x, \mathcal{I}, Q_{[N]}^{[x]})+ I(y_{[M]}; Q_{[N]}^{[x]}|x, \mathcal{I}) \\
 &=I(y_{[M]}; A_{[N]}^{[x]}|x, \mathcal{I}, Q_{[N]}^{[x]}) \label{eq_leakage_indep}\\
&= I(y_{[M]}; A_{[N]}^{[x]}|x, \mathcal{I}), \label{dummy}
\end{align}
where \eqref{eq_leakage_indep} follows since $y_{[M]}=(y_1,\ldots,y_M)$ are independent of the queries $Q_{[N]}^{[x]}$ given $x$ and $\mathcal{I}$, and \eqref{dummy} holds for both schemes because the randomness symbols in $Q_{[N]}^{[x]}$ added for the privacy of $x,\mathcal{I}$, upon answer generation are protected by the randomness shared among the servers. Additionally, they align on a subspace different from the database information subspace, leading to no leakage from the interference terms. In particular, the database leakage for the two-phase scheme originates from $\rho_i||h_1\circ(y_i-x)||^2$ in \eqref{answer_phase1} and $||h_2\circ y_i-x||^2$ in \eqref{phase2_ipcr_leakage}. Similarly, $(y_i-x)^T\big((y_i-x)\circ h\big)$ in \eqref{answer_single_phase} is responsible for leakage in the single-phase scheme.

For comparing the database leakage between the two schemes, define for each $i\in [M]$,
\begin{align}
    E_i = \mathbbm{1}\left[x_{\mathcal{I}}=y_{i,\mathcal{I}}\right].
\end{align}
Thus, the answers for both schemes satisfy
\begin{align}\label{eq_E_funcA}
     H(E_{[M]}|A_{[N]}^{[x]}, x, \mathcal{I})&=0,
\end{align}
where $E_{[M]}=(E_1,\ldots,E_M)$. The database leakage is
\begin{align}\label{eq_E_y_joint}
    &I(y_{[M]}; A_{[N]}^{[x]}|x,\mathcal{I})=I(E_{[M]},y_{[M]}; A_{[N]}^{[x]}|x, \mathcal{I}), 
\end{align}
since $I(E_{[M]};A_{[N]}^{[x]}|x,\mathcal{I}, y_{[M]})=0$. The right hand side of \eqref{eq_E_y_joint} can be expanded as
\begin{align}
     I(E_{[M]};A_{[N]}^{[x]}|x,\mathcal{I})+ I(y_{[M]};A_{[N]}^{[x]}|x,\mathcal{I}, E_{[M]})= H(E_{[M]}|x,\mathcal{I}) + I(y_{[M]};A_{[N]}^{[x]}|x,\mathcal{I},E_{[M]}), \label{eq_leakage_compare}
\end{align}
where \eqref{eq_leakage_compare} follows from \eqref{eq_E_funcA}. The first term in the leakage appears for both schemes since $\Theta$ is inevitably learned by the user while computing $\theta^*$. The second term, however, is different for the two schemes.

For the two-phase scheme, the second term in \eqref{eq_leakage_compare} is the information that the user infers from the squared norm values $||y_i-x||^2=||y_{i,\mathcal{M}}-x_{\mathcal{M}}||^2$ only for $E_i=1$, i.e., $i\in \Theta$. On the other hand, for the single-phase scheme, the second term contains additional information about the immutable features of $y_i, i\notin \Theta$. In particular, if one immutable feature does not match, then $(y_i-x)^T\big((y_i-x)\circ h\big)\in \mathcal{J}_1=[L:LR^2+R^2(d-|\mathcal{I}|)]$ whereas if two immutable features do not match, then $(y_i-x)^T\big((y_i-x)\circ h\big)\in \mathcal{J}_2=[2L:2LR^2+R^2(d-|\mathcal{I}|)]$ and so on. In general $\mathcal{J}_k=[kL:kL+R^2(d-|\mathcal{I}|)]$ is the set where $(y_i-x)^T\big((y_i-x)\circ h\big)$ lies if $k$ out of $|\mathcal{I}|$ features do not agree. These sets, though overlapping, leak non-zero amount of information on $y_i, i\notin \Theta$. Another way to see that the leakage in the single-phase scheme is higher, is because, the user observes the values of $(y_i-x)^T\big((y_i-x)\circ h\big)$ for $i\in [M]\setminus \Theta$, in addition to learning $E_{[M]}$. Whereas, for the two-phase scheme, no additional information about $y_i$ is learnt by the user if $i\in [M]\setminus \Theta$.
\begin{remark}
 From PCR to I-PCR, the number of replicated servers that participate for our schemes increases from $N=2$ to $N=3$. This is due to the added constraint of hiding the immutable set $\mathcal{I}$ from the institution, which necessitates an additional dimension to be resolved, requiring an extra server.
\end{remark}
\section{Incorporating User-Actionability}\label{sec:user_act}
In this section, we implement user-actionability, i.e., the user's preference to modify some features over others (among the ones they are willing to change), to obtain a more actionable counterfactual, while keeping their preference private from the institution. To model this, we assume that the user assigns different \emph{weights} to their attributes based on their preference to change those attributes to attain a counterfactual. The more reluctant the user is to change a given attribute, the higher is the weight of that attribute. Accordingly, the distance metric is now the weighted $\ell_2$ norm whose coordinates are scaled by the weight vector. In order to ensure that the user's preference is also private, the weight vector should be kept private from the servers.

\subsection{PCR+ Schemes}
Let $w\in [L_1]^{d}$ denote the user's weight vector, where $L_1$ is a positive integer. The modified distance metric is given by
\begin{align}\label{dist_pcr_weighted}
    d(x,y)=(y-x)^T\left((y-x)\circ w\right).
\end{align}
The goal is same as \eqref{goal_pcr} with $d(\cdot,\cdot)$ modified according to \eqref{dist_pcr_weighted}.  The decodability \eqref{eq:decodability_pcr}, user-privacy \eqref{eq:user_priv_pcr} and database leakage \eqref{eq:leakage_pcr} definitions are modified to include $w$ with the random variable $x$ in the arguments of $H(\cdot)$ and $I(\cdot;\cdot)$. Then, we have the following theorem.

\begin{theorem}\label{pcr_weighted_thm}
    There exist schemes derived from Baseline PCR, Diff-PCR and Mask-PCR, that can retrieve the exact closest counterfactual with user's actionability weights using $N = 3$ servers while keeping the user's sample and actionability weights hidden from each server. The communication costs of the respective schemes are:
    \begin{enumerate}
        \item Baseline PCR+: $6d+3M$ symbols of $\mathbb{F}_q$, $q>R^2 L_1d$
         \item Diff PCR+: $6d+3(M-1)$ symbols of $\mathbb{F}_q$, $q> 2R^2L_1d$
        \item Mask PCR+: $6d+3M$ symbols of $\mathbb{F}_q$, $q>R^2L_1d$
    \end{enumerate}
    where $q$ is prime. 
\end{theorem}

To prove this, we describe the extended schemes next. Since it is clear from the context, we use the same notations as the PCR setting to denote the queries and answers.

\subsubsection{Baseline PCR+}\label{baseline_weighted} 
For this scheme, the required minimum field size is $q > R^2L_1d$. Now, the user sends the query tuple $Q_n^{[x]}=[Q_n^{[x]}(1), Q_n^{[x]}(2)]$ to server $n$, where
\begin{align}\label{queries_weighted}
    Q_n^{[x]}(1) = x+\alpha_n Z_1, \quad Q_n^{[x]}(2) = w + \alpha_n Z_2,
\end{align}
and $Z_1$ and $Z_2$ are uniform independent random vectors in $\mathbb{F}_q^d$. Upon receiving the queries, the answers are generated as follows,
\begin{align}
   A_n^{[x]}(i) = &(y_i-Q_n^{[x]}(1))^T\diag(Q_n^{[x]}(2)) (y_i-Q_n^{[x]}(1))+\alpha_n Z'_1(i) + \alpha_n^2 Z'_2(i)\\
     =& d(y_i,x) + \alpha_n \Big( (y_i-x)^T \left((y_i-x)\circ Z_2\right) -2(y_i-x)^T\left(Z_1 \circ w\right) +Z'_1(i)\Big) \nonumber \\ &+\alpha_n^2 \Big( Z_1^T \left(Z_1 \circ w\right)- 2(y_i-x)^T \left(Z_1 \circ Z_2\right)+Z'_2(i) \Big)+ \alpha_n^3 Z_1^T\left(Z_1 \circ Z_2\right),
\end{align}
where recall that $\circ$ denotes element-wise multiplication, and $Z'_2(i)$ are uniform random variables shared by the servers and used to invoke the one-time pad theorem.

Upon receiving the answers from $N=3$ servers, the user decodes them  $i\in[M], n\in [3]$ as follows
\begin{align}
    \hat{A}_n^{[x]}(i) &= A_n^{[x]}(i) - \alpha_n^3 Z_1^T\left(Z_1\circ Z_2\right),
\end{align}
since $Z_2$ is available to the user. Next,
\begin{align}
   \begin{bmatrix}
        \hat{A}_1^{[x]}(i)\\
        \hat{A}_2^{[x]}(i)\\
        \hat{A}_3^{[x]}(i)
    \end{bmatrix} &= \bm{V}_3\begin{bmatrix}
        d(y_i,x)\\
         I_1(i)\\
         I_2(i)
    \end{bmatrix},
\end{align}
where
\begin{align}\label{interference_weighted}
    I_1(i) &= (y_i-x)^T \left((y_i-x)\circ Z_2\right)-2(y_i-x)^T\left(Z_1 \circ w\right) +Z'_1(i) \\
    I_2(i) &=  Z_1^T \left(Z_1 \circ w\right)- 2(y_i-x)^T \left(Z_1 \circ Z_2\right) +Z'_2(i). 
\end{align}
The upload cost in this scheme is $6d$, while the download cost is $3M$.

\subsubsection{Diff-PCR+} 
The queries are given by \eqref{queries_weighted}. Similar to Diff-PCR, the servers construct their answers so that the user decodes only the difference of distances, instead of the exact distances. This time, the field of operation is a prime $q> 2R^2L_1d$, since, for all $i,j\in [M]$, the absolute difference $|d(y_i,x) - d(y_j,x)|=|(y_i-x)^T \left((y_i-x)\circ w\right)-(y_j-x)^T \left((y_j-x)\circ w\right)|$ is at most $R^2L_1d$. Therefore, in $\mathbb{F}_q$
\begin{align}
    d(y_i,x) - d(y_j,x)\in 
    \begin{cases}
        \mathcal{H},     &\text{$y_j$ is closer to $x$},\\
        \mathcal{H}^c,   &\text{$y_i$ is closer to $x$}, 
    \end{cases}
\end{align}
where $\mathcal{H}=[0:R^2L_1d]$ and $\mathcal{H}^c = [R^2L_1d+1:q-1]$. The user sends the same queries as in \eqref{queries_weighted}. Let the servers share two common randomness vectors ${Z}'_1=[Z'_1(1)\ldots Z'_1(M-1)]^T$ and ${Z}'_2=[Z'_2(1)\ldots Z'_2(M-1)]^T$ each of length $M-1$ where each entry is a uniform random variable from $\mathbb{F}_q$. Server $n\in [3]$ constructs the following answer for each $i\in [M-1]$
\begin{align}
    A_n^{[x]}(i)=&(y_i-Q_n^{[x]}(1))^T\text{diag}(Q_n^{[x]}(2))(y_i-Q_n^{[x]}(1))\nonumber\\
    &-(y_{i+1}-Q_n^{[x]}(1))^T\text{diag}(Q_n^{[x]}(2))(y_{i+1}-Q_n^{[x]}(1))+ \alpha_n Z_1'(i)+ \alpha_n^2 Z'_2(i)\\
    =& d(y_i,x) - d(y_{i+1},x)+\alpha_n I'_1(i) + \alpha_n^2 I'_2(i)
\end{align}
where $I_1'(i)=(y_i-x)^T \left((y_i-x) \circ Z_2\right) - (y_{i+1}-x)^T \left((y_{i+1}-x)\circ Z_2\right) +2Z_1^T\left((y_i-y_{i+1})\circ w\right)+Z'_1(i)$ and $I'_2(i)= 2(y_{i+1}-y_i)^T \left(Z_1 \circ Z_2\right) +Z'_2(i)$ are the interference terms. The answers $A_n^{[x]}(i)$ of the three servers can be written as
\begin{align}
    \begin{bmatrix}
        A_1^{[x]}(i)\\
        A_2^{[x]}(i)\\
        A_3^{[x]}(i)
    \end{bmatrix}=\bm{V}_3
    \begin{bmatrix}
        d(y_i,x) - d(y_{i+1},x)\\
        I_1'(i)\\
        I_2'(i)
    \end{bmatrix}.
\end{align}
Therefore, the user recovers the $M-1$ differences, while discarding the interference terms. Now, with the $M-1$ differences, the user evaluates their counterfactual index $\theta^*$ following sequential comparisons similar to Algorithm \ref{alg_difference_distance}, the only difference being the range in line \ref{range_prox} is replaced by $\mathcal{H}^c$. The upload cost in this scheme is $6d$, while the download cost is $3(M-1)$.

\subsubsection{Mask-PCR+}
The operating field size is the same as that of Baseline PCR+, i.e., $q>R^2L_1d$. Upon receiving the queries, as given in \eqref{queries_weighted}, each server applies the following on each sample $y_i$ and sends to the user
\begin{align}
    A_n^{[x]}(i) =& \big(y_i-Q_n^{[x]}(1)\big)^T\diag\big(Q_n^{[x]}(2)\big)\big(y_i-Q_n^{[x]}(1)\big)+\mu(i) +\alpha_n Z'_1(i)+ \alpha_n^2 Z'_2(i)\\
    =& d(y_i,x)+ \mu(i)+ \alpha_n I_1(i) + \alpha_n^2 I_2(i) +\alpha_n^3 Z_1^T \left(Z_1\circ Z_2\right),
\end{align}
where $\mu(i)$ is as defined in Section \ref{masking_equal_actionable}, $Z'_1(i)$, and $Z'_2(i)$ are uniform random variables and $I_1(i)$ and $I_2(i)$ are as given in \eqref{interference_weighted}. Using the answers of the $N=3$ servers and applying the decoding approach given in Section~\ref{baseline_weighted}, the user obtains the masked weighted distance corresponding to $y_i$ as follows
\begin{align}
    \begin{bmatrix}
        \hat{A}_1^{[x]}(i)\\
        \hat{A}_2^{[x]}(i)\\
        \hat{A}_3^{[x]}(i)
    \end{bmatrix} =  \bm{V}_3\begin{bmatrix}
        d(y_i,x) +\mu(i)\\
         I_1(i)\\
         I_2(i)
    \end{bmatrix}.
\end{align}

To show that there is no need to change the range of the masking random variable $\mu$, under the modified distance metric, we provide the following lemma.

\begin{lemma}\label{lemma_range_fixed}
    The range of the random variable designed to mask the exact distance in this case is the same as in the non-weighted case.
\end{lemma}

The proof proceeds similarly as that of Lemma~\ref{Mask-PCR-decoding-lemma} and is relegated to Appendix~\ref{lemma_range_fixed_proof}. The upload cost in this scheme is $6d$, and the download cost is $3M$.

\subsection{I-PCR+ Schemes} 
Under the immutability constraint, the weight vector is restricted to the mutable features, i.e., those in $\mathcal{M}$. To realize this, the user assigns an integer $w'(k)\in [L_1]$ for each $k\in \mathcal{M}$, where a higher value indicates more reluctance to change that attribute. Then, the user's goal is to find the index $i$ that minimizes
\begin{align*}
    (y_{i,\mathcal{M}}-x_{\mathcal{M}})^T \big((y_{i,\mathcal{M}}-x_{\mathcal{M}})\circ [w'(k)]_{k\in \mathcal{M}}\big)
\end{align*}
over all $y_i\in \mathcal{D}$ such that $y_{i,\mathcal{I}}=x_{\mathcal{I}}$, i.e., $i\in \Theta$. The decodability \eqref{eq:ipcr_decodability}, user privacy \eqref{eq:ipcr_privacy} and \eqref{eq:ipcr_db_leakage} are modified to include $w$ as an additional argument in $H(\cdot)$ and $I(\cdot;\cdot)$, along with $x$ and $\mathcal{I}$. Then, the I-PCR schemes can be modified to achieve user-actionability as stated in the following theorem.

\begin{theorem}\label{ipcr_weighted_thm}
    There exist schemes derived from Two-Phase I-PCR and Single-Phase I-PCR that can retrieve the exact closest counterfactual under the immutability constraint with user's actionability, while keeping the user's sample and actionability weights hidden from each server. The communication costs of the respective schemes are:
    \begin{enumerate}
        \item Two-Phase I-PCR+: $14d+11M$ symbols of $\mathbb{F}_q$, $q> R^2L_1d$
         \item Single-Phase I-PCR+: $6d+3M$ symbols of $\mathbb{F}_q$, $q>F(L-1)R^2 + R^2L_1d$
    \end{enumerate}
    where $q$ is prime.  
\end{theorem}

\subsubsection{Two-Phase I-PCR+}
For the two-phase I-PCR scheme, the field size requirement is now $\mathbb{F}_q$ with $q>L_1R^2d$. The first phase proceeds in the same manner as that of the Two-Phase I-PCR scheme (Section \ref{subsec:ipcr_2phase}), and the user obtains $\Theta$. In the second phase, the user creates a vector $w$ whose $k$th entry is given by
\begin{align}
    w(k)=\begin{cases}
        1, & k\in \mathcal{I},\\
        w'(k), & k\in \mathcal{M}.
    \end{cases}
\end{align}
Then, the user sends the following query to server $n$
\begin{align}
    Q_n^{[x_{\mathcal{M}}]} = & \begin{bmatrix}Q_n^{[x_{\mathcal{M}}]}(1) &  Q_n^{[x_{\mathcal{M}}]}(2) & w+\alpha_n Z_5\end{bmatrix} \\ = & \begin{bmatrix}h_2 + \alpha_n Z_3 & x + \alpha_n Z_4 & w+\alpha_n Z_5\end{bmatrix},
\end{align}
where $h_2$ is defined in \eqref{define_h2}, $Z_3$, $Z_4$ and $Z_5$ are uniform vectors, chosen independently from $\mathbb{F}_{q}^{M}$, $\mathbb{F}_{q}^{d}$ and $\mathbb{F}_q^d$, respectively. Correspondingly, the answer returned by server $n$ for each $i\in [M]$ is
\begin{align}
    A_n^{[x_{\mathcal{M}}]} (i) = &\big(\tilde{S}_n (i)^T - Q_n^{[x_{\mathcal{M}}]}(2)\big)^T\big(\tilde{S}_n (i)^T - Q_n^{[x_{\mathcal{M}}]}(2) \circ (w+\alpha_n Z_5)\big)+ \alpha_n Z'_3(i) + \alpha_n^2 Z'_4(i) \nonumber\\
    &+\alpha_n^3 Z'_5(i)\\
    \label{eq:ipcr_2ph+_ans}=&  (h_2(i) y_{i} - x)^T\left((h_2(i) y_{i} - x)\circ w\right) + \alpha_n I_{31}(i) + \alpha_n^2I_{32}(i) + \alpha_n^3 I_{33}(i),
\end{align}
where $\tilde{S}_n(i)$ is the $i$th row of the masked database defined in \eqref{define_masked_db}, $Z_3'(i)$, $Z_4'(i)$ and $Z_5'(i)$ are uniform and independent random variables from $\mathbb{F}_q$. Since \eqref{eq:ipcr_2ph+_ans} is a polynomial of degree $3$ in $\alpha_n$, the user requires answers from $N=4$ servers to perform decoding as follows
\begin{align}
    \begin{bmatrix}
        A_1^{[x_{\mathcal{M}}]}(i)\\
        A_2^{[x_{\mathcal{M}}]}(i)\\
        A_3^{[x_{\mathcal{M}}]}(i)\\
        A_4^{[x_{\mathcal{M}}]}(i)
    \end{bmatrix} = \bm{V}_4
    \begin{bmatrix}
         (h_2(i) y_{i} - x)^T\left((h_2(i) y_{i} - x)\circ w\right)\\
        I_{31}(i)\\
        I_{32}(i)\\
        I_{33}(i)
    \end{bmatrix},
\end{align}
where the first term reduces to
\begin{align}
 (h_2(i) y_{i} - x)^T\left((h_2(i) y_{i} - x)\circ w\right)=
 \begin{cases}
     (y_{i} - x)^T\left((y_i-x)\circ w \right), & i\in \Theta,\\
     x^T(x\circ w), & \text{otherwise.}
 \end{cases}
\end{align} 
The upload and download costs in the first phase are $6d$ and $3M$, respectively, while those in the second phase are $8d+4M$ and $4M$, respectively. This results in the total communication cost of $14d+11M$ symbols.

\subsubsection{Single-Phase I-PCR+}
With the single-phase scheme, actionability can be directly implemented by modifying \eqref{weight_assignment} as follows
\begin{align}
    h(k) = \begin{cases}
        L, & k\in \mathcal{I},\\
        w'(k), & k\in \mathcal{M},
    \end{cases}
\end{align}
where $L>L_1R^2d$. The rest of the scheme follows the description in Section \ref{ipcr_single_phase} with $N=3$ servers. This leads to the same communication cost of $6d+3M$ symbols, as that of the single-phase scheme, however, in a larger field $\mathbb{F}_q$, with $q>F(L-1)R^2 + L_1R^2d$.

\begin{remark}
    Extensions to PCR+ from PCR, and to two-phase I-PCR+ from I-PCR, require an additional dimension to preserve the privacy of the user's weight vector. This is in addition to the requirement of a larger field size. Only in the single-phase I-PCR+ scheme, the number of servers, and the communication cost are not affected by actionability. However, this comes at the expense of a much higher field size. 
\end{remark}

\section{Experiments}\label{sec:expts}
In this section, we describe the experiments performed on both synthetic and real-world datasets. For each experiment, we describe the dataset being used, and how they are processed before implementation. The first experiment demonstrates the dependence of the accuracy of the Baseline PCR scheme on the operating field size, which is decided by the number of quantization levels $R$. Additionally, for the Mask-PCR scheme, we empirically determine $d_{\min}$ to lift the restriction that $x\in\mathcal{D}_c$. In our next experiment, we record the database leakages of the Baseline, Diff- and Mask-PCR schemes on a synthetic and real dataset, for various values of $d_{\min}$. Finally, we numerically evaluate the database leakage for the Two-Phase and Single-Phase I-PCR schemes, for various sizes of $\mathcal{I}$.
\subsection{Accuracy-Quantization Trade-Off in PCR}\label{exp:accu_quant}

\begin{figure}[h]
    \centering
    \includegraphics[width=0.6\textwidth]{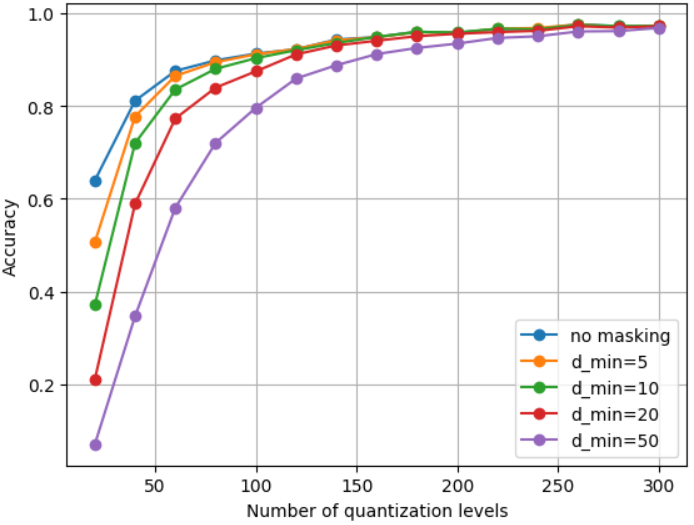}
    \caption{Accuracy-quantization trade-off.}
    \label{fig_accQuantizationTradeOff}
\end{figure}

As detailed in Section \ref{userpriv1}, the proposed schemes operate on finite fields with size $q$. However, a real-world application might require some of the features to be real-valued. To circumvent this, we quantize each feature of the counterfactual instances as well as the user sample before applying the schemes to align with our system model. However, the quantization of each feature to integers in $[0:R]$ will lead to a loss of some information that might result in an error in finding the closest counterfactual. The error can be made arbitrarily small by increasing the number of quantization levels. The greater the value of $R$, the larger is the field size $q$ for absolute accuracy.

Moreover, in the case of Mask-PCR, we lift the restriction $x\in\mathcal{D}_c$ off by empirically deciding the parameter $d_{\min}$ to be used such that the scheme works for most of the points in a sample set of $x$. As pointed out in Section \ref{masking_equal_actionable}, a larger field size $q$ is needed to accommodate a larger value for $d_{\min}$. 

\paragraph{Wine Quality Dataset:} The Wine Quality Dataset \cite{wineDataset} contains 4898 instances, each with 11 real-valued features. The original target variable ``Quality'' is categorical, taking integer values from 0 to 10. We convert the target into a binary variable by defining $t=\mathds{1}[``\text{Quality}"\geq 5]$. This gives us 3788 instances with $t=1$ and 183 instances with $t=0$ (which are the potential rejected instances).

We consider all the instances with $t=1$ to be the accepted (potential counterfactual) instances. Let this set be $\mathcal{S}_1$. The set $\mathcal{S}_0$ consists of all the rejected instances, i.e., with $t=0$. In each round, we pick a subset $\mathcal{D}$, with $|\mathcal{D}|=M$, of $\mathcal{S}_1$ as the database and a subset $\mathcal{S}_x$ of $\mathcal{S}_0$ as the rejected instances. Then, we apply the retrieval schemes with varying parameters to obtain the nearest neighbor counterfactuals. The accuracy of the retrieved counterfactuals is computed as follows: Denote the counterfactual of  $x$ retrieved using the scheme by $\hat{y}(x)$ and the actual (i.e., the exact nearest neighbor) counterfactual by $y(x)$. Note that the scheme provides the index $\theta^*$ of the counterfactual, hence, we can retrieve the exact unquantized counterfactual irrespective of the quantization applied. The accuracy of the scheme is  
\begin{align}
    \text{Accuracy} = \frac{1}{|\mathcal{S}_x|}  \sum_{x_i\in\mathcal{S}_x}   \mathds{1}\Big[||x_i-\hat{y}(x_i)||\leq ||x_i-y(x_i)||\Big].
\end{align}
We average the accuracy over several rounds with uniformly sampled $\mathcal{D}\subset\mathcal{S}_1$ and $\mathcal{S}_x\subset\mathcal{S}_0$. We set the database size $M$ to be 500 and the number of queries per round (i.e., $|\mathcal{S}_x|$) to be 50. We repeat the experiment for 100 rounds. 

Fig.~\ref{fig_accQuantizationTradeOff}, shows how the accuracy improves with an increasing number of quantization levels. Further, for a given number of quantization levels, the accuracy degrades as $d_{\min}$ increases. Note that, when $d_{\min}=1$, the masking scheme reduces to the Baseline PCR scheme, represented by the top blue curve.

\subsection{Database Leakage in PCR}\label{expt:leakage_pcr}
To observe the success of Diff-PCR and Mask-PCR in mitigating the leakage, and evaluate their exact leakage values, we proceed by following the steps in Appendix \ref{leakage_comp_app}. 

\paragraph{Synthetic Dataset:} 
The samples are assumed to be uniformly distributed over $[0:R]^d$ space. For each rejected instance $x$, the counterfactuals $y_1, \ldots, y_M$ are assumed to be uniformly distributed over $[0:R]^d \backslash \{x\}$. We consider $R=4$ and $d=2$ with $M=5$. We assume the queries are equi-probable over the $[0:R]^d$ space. Table \ref{table_leakageValues_pcr} lists the computed leakage values under these assumptions.

\paragraph{COMPAS Dataset:}
Correctional Offender Management Profiling for Alternative Sanctions (COMPAS) dataset \cite{compas} includes 6172 instances and 20 features (after one-hot encoding the categorical features) about individuals, which are used to predict the likelihood of reoffending of a convicted person. The target variable $t$ is ``is\textunderscore recid''. Class-wise counts are 3182 and 2990 for $t=0$ and $t=1$, respectively. We normalize all the features to the interval $[0,1]$ and quantize to $R+1$ levels with $R=10$ during pre-processing, leaving 560 unique instances with $t=0$ and 717 unique instances with $t=1$. As our schemes are independent of the classifier, we use these labels as the classifications of the instances. 

In order to compute the leakage values, we assume the following: 1) Queries $x$ are equi-probable over the $t=0$ (rejected) instances. Therefore, $\prob{X}{x}=\frac{1}{560}$ (denoted by $\rho_X$). 2) Since computing the histogram $\mathcal{C}(x)$ in \eqref{eq_leakageSimFinal} over all possible $M$-tuples (with $M=5$ in this particular experiment) is too computationally intensive, we sample 100000 of the possible permutations and evaluate the leakage over those. Accordingly, given the query $x$, the $M$-tuples $y=(y_1,\dots,y_M)$ are assumed to be equi-probable over a set of 100000 $M$-tuples that can be generated using the $t=1$ (accepted) instances. Therefore, $\prob{Y|X}{y|x}=\frac{1}{100000}$ (denoted by $\rho_{Y|X}$). Table~\ref{table_leakageValuesCompas} presents the results. Similar to the values obtained with the synthetic dataset, we observe that Diff-PCR and Mask-PCR schemes reduce leakage compared to the baseline scheme.

\begin{table}[h!]
    \centering
    \begin{tabular}{|c|c|} 
    \hline
    Scheme & Leakage (to base $q=757$) \\ [0.5ex] 
    \hline
    Baseline PCR & 3.1297 \\
    Diff-PCR & 2.0404 \\
    Mask-PCR ($d_\text{min}=2$) & 0.911 \\
    Mask-PCR ($d_\text{min}=3$) & 0.777 \\
    \hline
    \end{tabular}
    \caption{Leakage results for PCR schemes -- Synthetic dataset.}
    \label{table_leakageValues_pcr}
\end{table}

\begin{table}[htb]
    \centering
    \begin{tabular}{|c|c|} 
    \hline
    Scheme & Leakage (to base $q=2003$) \\ [0.5ex] 
    \hline
    Baseline PCR & 1.51437 \\
    Diff-PCR & 1.51436 \\
    Mask-PCR ($d_\text{min}=2$) &  1.51427\\
    \hline
    \end{tabular}
    \caption{Leakage results for PCR schemes -- COMPAS dataset.}
    \label{table_leakageValuesCompas}
\end{table}

\subsection{Database Leakage in I-PCR}
We compute the leakage values as follows. For the single-phase scheme, we compute the leakage as
\begin{align}
    I(y_{[M]};(y_i-x)^T\big((y_i-x)\circ h\big), \quad i\in [M]|x,\mathcal{I}).
\end{align}
For the two-phase scheme, we simplify the database leakage expression using \eqref{eq_E_y_joint} and \eqref{eq_leakage_compare}, to obtain,
\begin{align}
    I(y_{[M]}; A_{[N]}^{[x]}|x,\mathcal{I})=H(E_{[M]}|x,\mathcal{I})+I(y_{[M]};A_{[N]}^{[x]}|x,\mathcal{I},E_{[M]}).
\end{align}
Expanding the latter term gives,
\begin{align} \label{derive_sec_term}
    I(y_{[M]}; A_{[N]}^{[x_{\mathcal{I}}]}, A_{[N]}^{[x_{\mathcal{M}}]}|x,\mathcal{I},E_{[M]}) =
    I(y_{[M]}; A_{[N]}^{[x_{\mathcal{I}}]}|x, \mathcal{I},E_{[M]})+I(y_{[M]}; A_{[N]}^{[x_{\mathcal{M}}]}|x,\mathcal{I}, E_{[M]},A_{[N]}^{[x_{\mathcal{I}}]})
\end{align}
where the first term on the right hand side is $0$ since the answers after the first phase reveal no more information on $y_{[M]}$ beyond $E_{[M]}$. Now, since $H(A_{[N]}^{[x_{\mathcal{M}}]}|x,\mathcal{I},E_{[M]}, A_{[N]}^{[x_{\mathcal{I}}]},y_{[M]})=0$, we re-write the second term in \eqref{derive_sec_term} as
\begin{align}
    I(y_{[M]}; A_{[N]}^{[x_{\mathcal{M}}]}|x, \mathcal{I},E_{[M]},A_{[N]}^{[x_{\mathcal{I}}]})
    &=H(A_{[N]}^{[x_{\mathcal{M}}]}|x,\mathcal{I}, E_{[M]},A_{[N]}^{[x_{\mathcal{I}}]})\\
    &= H(A_{[N]}^{[x_{\mathcal{M}}]}|x,\mathcal{I}, E_{[M]}). \label{phase1_ans_indep}
\end{align}
Here, \eqref{phase1_ans_indep} follows because, for each $i\in[M]$, after decoding $A_{[N]}^{[x_{\mathcal{I}}]}(i)$, can be represented as
\begin{align}
    \rho_i||y_{i,\mathcal{I}}-x_{\mathcal{I}}||^2=\nu_i E_i,
\end{align}
for some $\nu_i$. Further, $I(A_{[N]}^{[x_{\mathcal{M}}]}; \nu_i E_i, i\in[M] | x, \mathcal{I}, E_{[M]}) = 0$ since $\rho_i$ is chosen at random from $[1:q-1]$ independently, and thus $\nu_i$ is also independent from $y_{[M]}$, $x$, and $Q_n^{[x_{\mathcal{M}}]}$, making it independent of $A_{[N]}^{[x_{\mathcal{M}}]}$.

Using the above simplification, we proceed to compute the exact leakage values on the synthetic dataset, setting $R = 3$, $d = 3$, and the database size $M = 3$.  Moreover, the set of immutable features $\mathcal{I}$ is assumed to be uniformly distributed over $[d]$, while having a fixed cardinality. For the single-phase scheme, $F=d$, $L=R^2d+1=28$ and $q$ is chosen as $757$ which is the smallest prime greater than $d(L-1)R^2+R^2d=756$. Table~\ref{table_leakageValues} presents the computed leakage values under these conditions, with varying $|\mathcal{I}|$ and logarithms taken to base $q = 757$ for both schemes. 

\begin{table}[h]
     \centering
     \begin{tabular}{|c |c | c|} 
     \hline
     $|\mathcal{I}|$ & Single-phase I-PCR & Two-phase I-PCR \\ 
     \hline
     0 & 1.1432 & 0.9422 \\
     1 & 1.4492 & 0.4200 \\
     2 & 1.4492 & 0.0977 \\
     3 & 1.1432 & 0.0000 \\
     \hline
     \end{tabular}
     \caption{Leakage results for I-PCR schemes with varying $|\mathcal{I}|$ to the base $q=757$.}
     \label{table_leakageValues}
\end{table}

As shown in Table~\ref{table_leakageValues}, the leakage for the two-phase scheme is always lower than that for the single-phase scheme. For the single-phase scheme, the leakage values are equal for $|\mathcal{I}|$ and $d-|\mathcal{I}|$. As opposed to the single-phase scheme, the leakage value of the two-phase scheme decreases as $|\mathcal{I}|$ increases. This is because, the expected size of $\Theta$ reduces, revealing less information on $\mathcal{D}$. The case of $|\mathcal{I}|=d$ is interesting since the user is guaranteed to not find a counterfactual. For the two-phase scheme, the leakage is $0$ since the user knows that $\Theta=\emptyset$ since $x$ is rejected, hence $x\notin \mathcal{D}$. However, this is non-zero for the single-phase scheme.

\section{Conclusion}\label{sec:conclude}
In this work, we investigated the problem of counterfactual retrieval, with information-theoretic privacy of the applicant against an institution. We started with formulating basic PCR, where the user is flexible to change any of their features to obtain the closest accepted sample. We proposed three schemes, which ensure varying levels of protection for the institution's database. We modified the PCR formulation to include the immutability constraint which fixes the values of certain features to not change in the counterfactual. Under this setup, we proposed two schemes with perfect user privacy, and varying levels of database privacy. Finally, we accounted for user actionability where the user can have different preference to change different attributes, and showed that our schemes can be extended to accommodate this setting. In addition to perfect user privacy, our schemes guarantee database privacy against the user, to some degree. Our experimental results support our findings on the comparative leakage values. Depending on the privacy budget, the institution can choose to follow one scheme over another.

\appendix
\section{Appendix}
\subsection{Baseline PCR With Other Metrics}\label{apend_high_norm}
If the metric for closeness is the $\ell_k$ norm with $k$ even, i.e., $d(x,y) = ||y-x||^k$, then the chosen $\mathbb{F}_q$ grows to have $q>R^kd$. The user sends the same query as \eqref{queries_no_weight}. Upon receiving the query, the servers compute their answer for $y_i\in \mathcal{D}$ as
\begin{align}
    A_n^{[x]}(i) = ||y_i-(x+\alpha_n Z)||^k + \sum_{\ell=1}^{k-1}\alpha_n^\ell Z'_\ell.
\end{align}
Correspondingly, the user decodes the answers to first compute
\begin{align}
    \hat{A}_n^{[x]}(i) = A_n^{[x]}(i) - \alpha_n ^k Z^k \mathbf{1}_k
\end{align} 
where $\mathbf{1}_k$ is the all-ones column vector, and obtain $d(y_i,x)$ since
\begin{align}
    \begin{bmatrix}
       \hat{A}_1^{[x]}(i)\\ 
       \hat{A}_2^{[x]}(i)\\ 
       \vdots\\
       \hat{A}_k^{[x]}(i)\\ 
    \end{bmatrix}
    =\bm{V}_k 
    \begin{bmatrix}
         d(y_i,x)\\
         I_1(i)\\
         \vdots\\
         I_{k-1}(i)
    \end{bmatrix}
\end{align}
and $\bm{V}_k$ is invertible. Here, $I_\ell(i)$, $\ell\in [k-1], i\in [M]$ is the interference term
\begin{align}
    I_\ell(i) = (-1)^{\ell-1}\binom{k}{\ell}(y_i-x)^{k-\ell}Z^{\ell} + Z'_\ell .
\end{align}
The number of servers required for decodability is $N=k$.
For the dot product metric, i.e., $d(x,y) = y^T x$, our scheme requires $N=2$ and $q>R^2d$. The answer returned by server $n$ is
\begin{align}
    A_n^{[x]}(i) = y_i^T(x+\alpha_n Z) + \alpha_n Z',
\end{align}
and the decodability at the user follows since
\begin{align}
    \begin{bmatrix}
        A_1^{[x]}(i)\\
        A_2^{[x]}(i)
    \end{bmatrix}
    = \bm{V}_2 
    \begin{bmatrix}
        y_i^T x \\
        y_i^TZ+Z'
    \end{bmatrix}.
\end{align}
The user assigns $\theta^*$ to the $i\in[M]$ that maximizes $y_i^Tx$.
\subsection{Proof of Lemma \ref{Mask-PCR-decoding-lemma}}\label{lemma_mask_pcr_proof}
\begin{Proof}
    We want to show that if $x\in \mathcal{D}_c$, then, decodability can be ensured from the masked distances. First, assume that $x \in \{x_1,\ldots, x_K\}$. To prove the required result, it suffices to prove that the random variable used preserves the order of the distances between the samples, i.e., $d_i(x) - d_j(x) > 0$ if $y_j$ is closer to $x$ and $d_i(x) - d_j(x) < 0$ if $y_i$ is closer to $x$. Let $|d_{i}(x)-d_{j}(x)|=\ell$. Then,
    \begin{align}
         d_{i}(x)-d_{j}(x)= \begin{cases}
            \ell, & \text{$y_j$ is closer to $x$},\\ 
            -\ell, & \text{$y_i$ is closer to $x$}.
        \end{cases}
    \end{align}
    Now, assume $y_j$ is closer, thus
    \begin{align}
        d_{i}(x)-d_{j}(x) +\mu(i)-\mu(j) 
         &=\ell  +\mu(i)-\mu(j) \\
        & \geq \ell -(d_{\min}-1)\\
        & > 0,
    \end{align}
     and if $y_i$ is closer, we have
     \begin{align}
        d_{i}(x)-d_{j}(x) +\mu(i)-\mu(j)
         &=-\ell  +\mu(i)-\mu(j)\\
        & \leq - \ell +(d_{\min}-1)\\
        & < 0.
    \end{align}
    Now, assume $x \in \mathcal{D}_c$. Using the same approach as above, and the definition of $x \in \mathcal{D}_c\setminus\{x_1,\ldots,x_K\}$, we have $|d_i(x) - d_j(x)| \geq d_{\min}$, which proves the required result.
\end{Proof}

\subsection{Proof of Lemma~\ref{lemma_range_fixed}}\label{lemma_range_fixed_proof}
\begin{Proof}  
    Note that
    \begin{align}
      |(y_i-x_k)^T & \left((y_i-x_k)\circ w\right) -(y_j-x_k)^T\left((y_j-x_k)\circ w\right)| \nonumber \\
      & = \left| \sum_{\ell}   w_{\ell}(y_i(\ell)-x_k(\ell))^2 -w_{\ell}(y_j(\ell)-x_k(\ell))^2 \right|\\
      & \geq  \left|  \sum_{\ell}   (y_i(\ell)-x_k(\ell))^2 -(y_j(\ell)-x_k(\ell))^2 \right|\\
      &=  \left|d(y_i,x_k) - d(y_j,x_k)\right|.
    \end{align}
    Thus, $\min_{k} \min_{i,j} |(y_i-x_k)^T\left((y_i-x_k)\circ w\right) -(y_j-x_k)^T\left((y_j-x_k)\circ w\right)| \geq  d_{\min}$.
\end{Proof}

\subsection{Leakage Computation}\label{leakage_comp_app}
To compute the values in Table \ref{table_leakageValues_pcr} we make the following assumptions: 1) Rejected instance $x$ are equi-probable over the $[0:R]^d$ space. Therefore, $\prob{X}{x}=\frac{1}{(R+1)^d}$ (denote by $\rho_X$). 2) Given the instance $x$, the $M$-tuples $y=(y_1,\dots,y_M)$ are equi-probable over the $[0:R]^d\backslash \{x\}$ space. Therefore, $\prob{Y|X}{y|x}=\frac{1}{\;_{(R+1)^d - 1}P_M}$ (denote by $\rho_{Y|X}$) where $\;_aP_b = \frac{a!}{(a-b)!}$ is the number of permutations.

Here, we present the approach we have taken to compute the leakage of the Baseline PCR scheme. A similar approach can be used to compute the values for Diff-PCR and Mask-PCR. Let $d(x)=(d_1(x),\dots,d_M(x))$. With the first assumption above, we simplify the leakage as follows:
\begin{align}
    I(y_1,\dots,y_M &;d_1(x),\dots,d_M(x)|x)\nonumber\\
    &= H(d_1(x),\dots,d_M(x)|x) - \underbrace{H(d_1(x),\dots,d_M(x)|x, y_1,\dots,y_M)}_{=0} \\
    &= H(d_1(x),\dots,d_M(x)|x) \\
    \label{eq_leakageSimBasic}
    &= -\sum_{x\in\mathcal{X}} \sum_{d(x)\in \mathcal{D}(x)} \prob{D(X),X}{d(x),x} \log(\prob{D(X)|X}{d(x)|x}) \\
    &= -\sum_{x\in\mathcal{X}} \sum_{d(x)\in \mathcal{D}(x)} \prob{D(X)|X}{d(x)|x} \prob{X}{x} \log(\prob{D(X)|X}{d(x)|x}) \\
    &= -\rho_X \sum_{x\in\mathcal{X}} \sum_{d(x)\in \mathcal{D}(x)} \prob{D(X)|X}{d(x)|x} \log(\prob{D(X)|X}{d(x)|x}).
\end{align}
Now, we focus on computing the conditional pmf $\prob{D(X)|X}{d(x)|x}$. Note that
\begin{align}
    \prob{D(X)|X}{d(x)|x} &= \sum_{y\in\mathcal{Y}} \prob{D(X), Y|X}{d(x), y|x} \\
    &= \sum_{y\in\mathcal{Y}} \prob{D(X)|Y, X}{d(x)| y, x} \prob{Y|X}{y|x} \\
    &= \rho_{Y|X}  \sum_{y\in\mathcal{Y}} \prob{D(X)|Y, X}{d(x)| y, x} \label{comment_added}\\
    &= \rho_{Y|X}  \sum_{y\in\mathcal{Y}} \prob{D(X)|Y, X}{d(x)| y, x} \\
    \label{eq_leakageSimConditionalPmf}
    &= \rho_{Y|X}  \sum_{y\in\mathcal{Y}} \mathds{1}\left[d(x)=(||y_1-x||^2,\dots,||y_M-x||^2)\right],
\end{align}
where \eqref{comment_added} is due to the second assumption above. Moreover, when summed over $\mathcal{D}(x)$ which is the set of all possible values for $d(x)$, \eqref{eq_leakageSimConditionalPmf} can be written as
\begin{align}
    \sum_{d(x)\in \mathcal{D}(x)} \prob{D(X)|X}{d(x)|x} &= \rho_{Y|X} \sum_{d(x)\in \mathcal{D}(x)}  \sum_{y\in\mathcal{Y}} \mathds{1}\left[d(x)=(||y_1-x||^2,\dots,||y_M-x||^2)\right] \\
    &= \rho_{Y|X} \underbrace{\sum_{y\in\mathcal{Y}} \sum_{d(x)\in \mathcal{D}(x)} \mathds{1}\left[d(x)=(||y_1-x||^2,\dots,||y_M-x||^2)\right]}_{=\mathcal{C}(x)}
\end{align}
where $\mathcal{C}(x)$ in the last term is the histogram of $d(x)$ over all $y$s for a given $x$. We compute this histogram by counting the $M$-tuples $(||y_1-x||^2, \dots, ||y_M-x||^2)$ while iterating over all possible $M$-tuples $y=(y_1,\dots,y_M)$ for the given $x$. Next we substitute this result back into \eqref{eq_leakageSimConditionalPmf} to obtain
\begin{align} \label{eq_leakageSimFinal}
    I(y_1,\dots,y_M;d_1(x),\dots,d_M(x)|x) = -\rho_X \rho_{Y|X} \sum_{x\in\mathcal{X}} \mathcal{C}(x) \log(\rho_{Y|X}\mathcal{C}(x)).
\end{align}

\bibliographystyle{unsrt}
\bibliography{references}

\end{document}